\documentclass[final,5p,times,twocolumn]{elsarticle}

\usepackage{amssymb}
\usepackage{amsmath}
\usepackage{url}
\usepackage{framed}
\usepackage{booktabs}
\usepackage{longtable}

\usepackage{tabularx}
\usepackage{graphicx}
\usepackage{arydshln}
\usepackage{makecell}
\usepackage{colortbl}
\usepackage{longtable}
\usepackage{multirow}
\usepackage{framed}
\usepackage[most]{tcolorbox}
\usepackage{amsfonts}
\usepackage{array}
\usepackage[table,xcdraw]{xcolor}
\usepackage{fancyhdr}
\usepackage{orcidlink}
\usepackage{subcaption}
\usepackage{url}
\usepackage{hyperref}
\hypersetup{hidelinks}
\usepackage{rotating}
\usepackage{longtable}

\journal{Journal of Systems and Software}

\begin{document}

\begin{frontmatter}



\title{A Cartography of Open Collaboration in Open Source AI: Mapping Practices, Motivations, and Governance in 14 Open Large Language Model Projects}


\author[1]{Johan Linåker}\corref{cor1}
\ead{johan.linaker@ri.se}
\ead[url]{https://orcid.org/0000-0001-9851-1404}

\author[2]{Cailean Osborne}
\ead{caileanosborne@gmail.com}
\ead[url]{https://orcid.org/0000-0002-4018-8488}

\author[3]{Jennifer Ding}
\ead{jending12@gmail.com}
\ead[url]{https://orcid.org/0000-0002-8266-4358}

\author[4]{Ben Burtenshaw}
\ead{benjamin.burtenshaw@huggingface.co}
\ead[url]{https://orcid.org/0009-0003-1276-7153}

\cortext[cor1]{Corresponding author.}

\address[1]{RISE Research Institutes of Sweden AB, Lund, Sweden}
\address[2]{University of Oxford, Oxford, UK}
\address[3]{Boundary Object Studio, London, UK}
\address[4]{Hugging Face, Antwerp, Belgium}




\begin{abstract}
The proliferation of open large language models (LLMs) is fostering a vibrant ecosystem in artificial intelligence (AI). However, the methods of collaboration used to develop open LLMs, both before and after their public release, have not yet been systematically studied, limiting our understanding of how open LLM projects are initiated, organised, and governed, as well as the opportunities to further foster this ecosystem. 
We address this gap through an exploratory analysis of open collaboration throughout the development and reuse lifecycle of open LLMs, drawing on semi-structured interviews with the developers of 14 diverse open LLM projects. 
These collaborations span multiple artefact domains---including models, data, software, evaluation, compute, and community engagement---each enabling distinct forms of participation and involving different stakeholders that evolves across the LLM development lifecycle, shifting from concentrated, selective engagement in the early stages to broader, distributed participation after model release. 
The open LLM developers are motivated by a variety of social, economic, and technological motivations, ranging from democratising access to AI and promoting open science to building regional ecosystems and expanding language representation.
These dynamics are coordinated through a range of governance structures, typically formal and professionalised to varying degrees, including centralised company-led efforts to decentralised grassroots initiatives. 
We synthesise our findings in a conceptual model of open collaboration in open LLM ecosystems, provide recommendations for practice, and conclude that openness in open source AI is not a uniform property but an emergent outcome of how collaboration is organised across interconnected artefact domains, lifecycle stages, and institutional contexts.
\end{abstract}




\end{frontmatter}




\section{Introduction}
\noindent The culture and practices of open collaboration in artificial intelligence (AI) research and development (R\&D) have evolved significantly in the last 20 years. In 2007, 16 eminent AI scientists came together to raise awareness that, while the machine learning (ML) research community had produced many powerful algorithms, their true potential remained unrealised because implementations were not openly shared~\cite{sonnenburg_need_2007}. The 2010s saw the emergence of open source software (OSS) ecosystems that have become key pillars of AI R\&D~\cite{langenkamp_how_2022}, in particular, foundational ML libraries such as scikit-learn~\cite{pedregosa_scikit-learn_2011} and deep learning frameworks such as PyTorch~\cite{paszke_pytorch_2019}. Platforms like arXiv became widely used to rapidly disseminate AI research~\cite{arxiv_arxivorg_2024}, and initiatives such as CommonCrawl~\cite{commoncrawl_common_2024} and ImageNet~\cite{imagenet_imagenet_2024} facilitated collaborations on datasets for training AI models. While the open release of foundational ML models, such as BERT~\cite{devlin_bert_2019} and YOLO~\cite{redmon_you_2016}, led to a rise in adoption and application of these models and the growth of the fields of Natural Language Processing and Computer Vision, respectively, the development and adoption of open AI models in this period remained limited.

Nowadays, AI R\&D thrives on the activity of global practitioners from many different kinds of organisations, who engage to share knowledge and collaborate on the development and maintenance of various projects and artefacts in the wider open source AI ecosystem, spanning OSS, open datasets, open models, open science projects, open standards, and open hardware~\cite{white_model_2024}. In particular, the widespread proliferation of open models---that is, AI models where the model parameters (i.e., weights and biases) and model architecture are publicly released under open source licenses along with model documentation~\cite{white_model_2024}---is fostering a vibrant ecosystem of collaborative research and innovation in AI, with almost two million models, including many powerful large language models (LLMs), now hosted on Hugging Face Hub~\cite{huggingface_models_2024}. With a growing community of millions of practitioners around the world releasing, reusing, and remixing millions of artefacts daily, progress on open LLMs is accelerating and branching into diverse application areas for domain and regional relevance. 

Researchers have begun studying developer practices in the burgeoning open source AI ecosystem, yielding four key insights for this study. First, despite the proliferation of open models, their reuse is highly concentrated; for example, only 1\% of models on the Hugging Face Hub account for 99\% of all downloads~\cite{osborne_ai_2024}. Second, developers on Hugging Face Hub exhibit ``nomadic'' engagement patterns, engaging intensely with models upon their release but quickly migrating to newer models and focusing on model applications rather than sustained collaboration~\cite{choksi_brief_2025}. Third, model maintenance activity tends to focus on ``perfective tasks'' such as performance enhancements, marking a key difference from typical OSS maintenance activities, such as bug fixing or feature development~\cite{castano_analyzing_2024}. Fourth, while most open LLMs follow a ``develop-then-release'' approach with minimal collaboration before the release, grassroots projects like EleutherAI~\cite{eleutherai_eleutherai_2021}, the BigScience Workshop~\cite{ding_towards_2023,akiki_bigscience_2022}, and Marin~\cite{hall2025marin} have pioneered community-led development of open LLMs (e.g., Pythia, BLOOM) and related artefacts, including open datasets (e.g., ROOTS) and open source LLM training and evaluation frameworks (e.g., GPT-NeoX, LM Evaluation Harness, lighteval).

Beyond these insights, we still have a limited understanding of the open collaboration methods that are used throughout the lifecycle of open LLMs, from the earliest pre-training stages through to downstream reuse and derivative development by the wider community. This gap limits our ability to effectively foster the open source AI ecosystem, such as by facilitating participation, connecting related projects, recommending or improving governance frameworks, and providing targeted support for under-resourced areas. We address this gap in our knowledge through an exploratory analysis of open collaboration on open LLMs, guided by the following research questions (RQs):


\begin{itemize}
    \item \textbf{(RQ1)} Where and how does open collaboration take place across the open LLM lifecycle?
    \item \textbf{(RQ2)} What incentivises engagement in the collaborative development of open LLMs?
    \item \textbf{(RQ3)} How are open collaborations on open LLMs coordinated and governed?
\end{itemize}   
 

To answer these RQs, we conducted semi-structured interviews with 17 developers from 14 open LLM projects, which represented diverse organisational contexts (grassroots projects, research institutes, startups, Big Tech companies) and geographic regions (North America, Europe, Africa, Asia), thus providing a comprehensive perspective on emerging approaches to open LLM collaboration around the world. 

In addressing the RQs, we make the following contributions:
\begin{itemize}
    \item We provide one of the first comprehensive qualitative studies of open collaboration practices across the full development and reuse lifecycle of open LLMs, based on interviews with developers from diverse organisational and geographical contexts.
    \item We identify and characterise six artefact domains—models, data, software, evaluation, compute, and community—showing how collaboration is distributed across interdependent artefacts rather than centred on models alone.
    \item We demonstrate how differences in participation access, expertise requirements, modularity, and control shape who can contribute and how collaboration unfolds across domains.
    \item We show how collaboration evolves across development stages, shifting from selective, centralised engagement in early phases to broader, distributed participation after release.
    \item We synthesise these findings into a conceptual model that links artefact domains, domain properties, lifecycle dynamics, motivations, and governance structures, explaining openness as an emergent property of open source AI ecosystems.
    \item We derive a set of recommendations for designing, organising, and supporting open collaboration in open source AI ecosystems.
\end{itemize}


The paper is structured as follows. In \S\ref{sec:llm-lifecycle}, we define key terms. In \S\ref{sec:relwork}, we survey prior work on open collaboration in AI R\&D. In \S\ref{sec:methods}, we present our research design. In \S\ref{sec:results-collabs}-\S\ref{sec:results-governance}, we present the findings on collaboration on-ramps and challenges throughout the open LLM lifecycle (RQ1), developer motivations (RQ2), and collaboration and governance frameworks in open LLM projects (RQ3). In \S\ref{sec:discussion}, we provide recommendations for diverse stakeholders seeking to support the global community of practitioners building a more open future for AI. In \S\ref{sec:conc}, we conclude the paper.

\begin{sidewaysfigure*}
    \centering
    \includegraphics[width=1\linewidth]{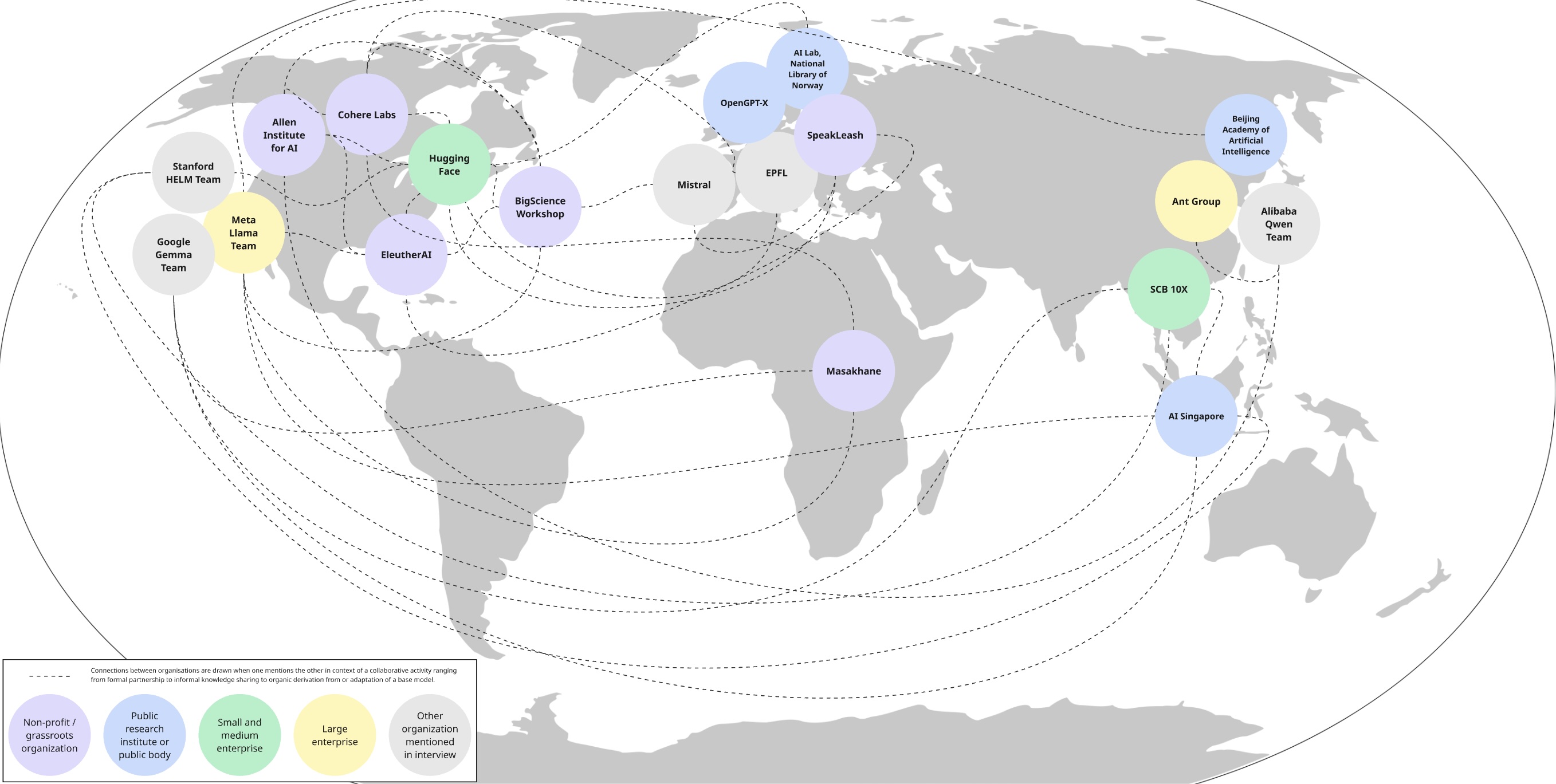}
    \caption{Cartography of Open Collaboration in Open Source AI across 14 Open LLM Initiatives}
    \label{fig:cartography}
\end{sidewaysfigure*}

 \section{Key terms}\label{sec:llm-lifecycle}

\subsection{LLMs}
\noindent LLMs are deep learning models, typically based on transformer architectures~\cite{vaswani_attention_2017}, that contain billions to trillions of parameters and are trained on extensive text corpora to learn statistical patterns in language, enabling them to generate coherent text, answer questions, and perform various language tasks without task-specific training. The term ``large language model'' was popularised by Brown et al.\ from OpenAI in their GPT-3 paper~\cite{brown_language_2020}.

\subsection{Open Collaboration}
\noindent We define ``open collaboration'' as a \textit{collaborative and coordinated production process between a number of individuals, where a publicly available common is used as input, and the output from the process is released back to the originating common}~\cite{linaaker2022sustaining, benkler2005common}. A common term in our context refers to an open LLM project and its various artefacts. This definition encompasses both direct collaborations (i.e., where two or more parties work together on a common goal) and indirect collaborations (i.e., development facilitated by open sharing, consisting mainly of reuse).


\subsection{Open LLM development and reuse lifecycle}
\noindent  We define the development and reuse lifecycle of open LLMs (henceforth: open LLM lifecycle) as the complete process from the beginning of the activities required to train an LLM through to how it is reused and redistributed after being publicly released in a repository on a platform like Hugging Face Hub, GitHub, or Modelscope. Our definition builds on Ding et al.'s~\cite{ding_2024_pipeline} ML model pipeline framework 
, which maps the technical pipeline of developing ML models, encompassing data collection, model training, and model deployment, alongside governance considerations that determine the possibility for input from external stakeholders (i.e., ``collaboration on-ramps''). We extend this framework to consider not only the pre-release development pipeline of LLMs but also post-release development and reuse activities, such as model fine-tuning and the collection of community feedback, thereby providing a more comprehensive view of open collaboration that may occur throughout the lifecycle of an open LLM. In particular, we focus on three stages in this lifecycle: pre-training, post-training, and post-release reuse.

\subsubsection{Pre-training}
\noindent  Model pre-training is the initial and most resource-intensive stage in the development of LLMs. It involves training a model from scratch on massive amounts of text data to learn general language understanding and generation capabilities. This foundational training allows the model to grasp grammar, context, and a range of knowledge. Key inputs for this stage 
include datasets, compute resources, the underlying software frameworks, and the specific model architecture design and implementation. The outputs of pre-training are model checkpoints, which are snapshots of the trained model at its best evaluation state, and initial evaluations to assess its performance on benchmark tasks. The process often involves defining the model's goals and requirements, followed by collecting and preparing pre-training data, ensuring its quality and relevance to the model's intended capabilities. The resulting output of this process is often referred to as a \textit{base model}, which is the term we will use in this paper.


\subsubsection{Post-training}
\noindent  Model post-training is the refinement stage that focuses on making a pre-trained (base) model specialised, aligned, or deployment-ready before being ``officially'' released. This pre-release stage typically involves techniques, such as supervised fine-tuning (SFT) using curated instruction-response pairs to teach the model to follow specific formats and tasks; reinforcement learning from human feedback (RLHF) or constitutional AI methods to align the model with human preferences and values; and safety filtering or guardrails to reduce harmful outputs. 
Key inputs include the pre-trained model checkpoints, supervised datasets, human preference data, and compute for iterative training. The process can also incorporate human annotators who provide feedback on model outputs, enabling the model to learn desired behaviours and response styles. Post-training outputs include refined model weights, safety evaluations, and performance benchmarks across various tasks. 


\subsubsection{Post-release reuse}
\noindent  Open LLMs are uploaded to platforms like Hugging Face Hub, GitHub, or Modelscope, and once publicly available, they can be reused in a myriad of ways, including deployment across different environments, fine-tuning for specialised languages or domains, quantisation to reduce computational requirements, knowledge distillation to create smaller models, and architectural modifications for specific hardware constraints, among others. This often involves taking a base model and making targeted modifications (e.g., finetuning, quantising, distilling models) to create a \textit{derivative model}, which may then be shared with the community by uploading it to repositories on the aforementioned platforms. This open ecosystem also facilitates community feedback, as users can share performance insights, identify failure modes, and suggest improvement priorities based on real-world deployment experience. Such feedback can reach base model developers through multiple channels, including GitHub issues for technical problems, discussions in Hugging Face Hub model repositories, and community communications via Discord, Slack, Reddit, or mailing lists.


\section{Related work}\label{sec:relwork}

\subsection{Open collaboration in AI R\&D}

\noindent  The practices and culture of open collaboration in AI R\&D have undergone significant shifts in the last two decades. While researchers lamented the scarcity of OSS for ML in 2007~\cite{sonnenburg_need_2007}, nowadays AI R\&D is hardly imaginable without open source ecosystems~\cite{langenkamp_how_2022} and collaboration in grassroots communities like scikit-learn~\cite{osborne_open_2024, pedregosa_scikit-learn_2011}, foundation-hosted ecosystems the LF AI \& Data Foundation and PyTorch Foundation~\cite{osborne_why_2024}, as well as company-led ecosystems like PyTorch (prior to its donation to the Linux Foundation) and TensorFlow~\cite{osborne_characterising_2025}. Beyond OSS, open collaboration in AI R\&D encompasses collaboration on open datasets, open science practices, open standards, open hardware, and increasingly open models~\cite{white_model_2024}. For example, platforms like arXiv are widely used to rapidly disseminate cutting-edge AI research~\cite{arxiv_arxivorg_2024}, and initiatives like CommonCrawl~\cite{commoncrawl_common_2024} and ImageNet~\cite{imagenet_imagenet_2024} maintain datasets for training AI models. Recently, open source AI practitioners have called for building an open human feedback ecosystem to facilitate AI alignment research, oriented around the sharing and collaborative annotation of user feedback to AI systems \cite{don-yehiya_future_2025}.

After years of debate about whether and how generative AI models, in particular LLMs, can be openly and responsibly released~\cite{solaiman_release_2019}, since 2021, the practice of making LLM weights publicly available in open repositories has become a popular choice for AI practitioners. In particular, model announcements or research paper publications are often accompanied by the model weights and other artefacts, similar to the former ``Papers with Code'' initiative~\cite{kang_papers_2023}. As a result, there are around 500,000 datasets and almost 2 million models on the Hugging Face Hub, which has emerged as a primary platform for practitioners to share, discuss, download, and build with models and datasets~\cite{gorwa_moderating_2024, laufer_anatomy_2025}. 

A growing corpus of research sheds light on the social dynamics of developer activity on Hugging Face Hub. While the number of models and datasets on the Hub is growing rapidly, activity on the platform is concentrated among a small fraction of model repositories and developers, and the vast majority of models receive limited contributions and downloads~\cite{osborne_ai_2024,choksi_brief_2025}. For example, 1\% of models account for 99\% of downloads~\cite{osborne_ai_2024}. Developer communities on the Hub exhibit distinct ``nomadic'' engagement patterns, where users engage intensely with models upon release but then migrate to newer models, creating successive activity spikes across different projects rather than sustained collaboration around individual models~\cite{choksi_brief_2025}. This nomadic behaviour fundamentally differs from traditional OSS development, where communities typically form around stable repositories with long-term governance structures~\cite{choksi_brief_2025}. In addition, Castaño et al. find that development activity in model repositories on the Hub differs from typical activity in OSS repositories on platforms like GitHub, with contributions to model repositories are typically oriented around ``perfective tasks'' that seek to improve model performance (e.g., optimization, fine-tuning, and evaluation) rather than corrective and adaptive tasks (e.g., bug fixes, feature additions).~\cite{castano_analyzing_2024}. Model repositories tend to have ``user communities'' rather than developer communities, where users engage with models as tools for applications rather than as artefacts to be collaboratively developed or extended~\cite{choksi_brief_2025}. These user communities often focus on requests for documentation, support, or expanded access rather than direct contributions to model improvement~\cite{choksi_brief_2025}.

While the majority of collaboration occurs post-release (e.g., fine-tuning, evaluation, or application development), a number of grassroots projects have pioneered community-driven collaboration from the earliest stages of LLM development. The BigScience Workshop represents one of the most ambitious examples, engaging over 1,000 contributors from 66 nations and 250 organisations in a coordinated effort to collaboratively develop the BLOOM model~\cite{akiki_bigscience_2022}. The project was coordinated by Hugging Face and benefited from French government support via subsidised access to the Jean-Zay supercomputer. It has been described as a ``values-driven'' initiative, which engaged contributors with diverse motivations, such as developing skills, publishing research, and contributing to the ecosystem~\cite{ding_towards_2023,akiki_bigscience_2022}. Other notable examples include EleutherAI's community-driven approach to developing resources,, such as evaluation frameworks like the Open LLM Leaderboard, which facilitate community-driven benchmarking of LLMs~\cite{biderman_lessons_2024}. Building on these precedents, newer projects like Marin are experimenting with even more radical forms of openness by creating an ``open labs'' that makes the entire research and development process of foundation models transparent from day one, leveraging established OSS development practices like GitHub-based workflows, pull requests, and community code review for foundation model development \cite{hall2025marin}

The collaborative trend in open source AI has accelerated with the 2024-2025 period marking unprecedented releases of high-performing models, including DeepSeek's V3 series, achieving competitive performance with proprietary models~\cite{deepseek-ai_deepseek-r1_2025}, Alibaba's Qwen 2.5 extending context windows to 1 million tokens~\cite{team_qwen2_2024}, and Google's Gemma 3 demonstrating efficient attention mechanisms~\cite{team_gemma_2025}. In August 2025, OpenAI, which had long opposed open-sourcing its industry-leading models, released its gpt-oss-120b and gpt-oss-20b models under an Apache 2.0 license \cite{openai_gpt-oss-120b_2025}.

\subsection{Openness vs. Open Source in AI}
\noindent  The proliferation of open models has generated much debate about definitions, as terms like ``open'' and ``open source'' are used inconsistently to describe AI models and \textit{``systems that offer minimal transparency or reusability…alongside those that offer maximal transparency, reusability, and extensibility''}~\cite{widder_open_2023}. In addition, the community wrestles with ``open-washing,'' where developers market models released under restrictive licenses as ``open source,'' which further confuses matters~\cite {liesenfeld_rethinking_2024,white_model_2024}. Several frameworks have been developed to clarify openness in AI. 

In terms of AI systems, the Open Source Initiative released the Open Source AI Definition version 1.0, which requires that an AI system and its constituent parts---i.e., model weights, code, and training data or sufficiently detailed information of it---are released under open source licenses for it to qualify as ``open source AI''~\cite{osi_open_2024}. Taking a different approach, Irene Solaiman proposes that the openness of generative AI systems can be plotted along a six-tiered gradient, from fully open to fully closed, involving distinct release methods and cost-benefit considerations~\cite{solaiman_gradient_2023}. Similarly, Basdevant et al. propose that openness should be understood across the stack of AI systems, spanning infrastructure, model components, and product/UX ~\cite{basdevant_towards_2024}. Meanwhile, the Digital Public Goods Alliance's definition of open source AI adds requirements on documented evidence of platform and technology neutrality, as well as ethical use cases~\cite{dpga2024core}. 

In terms of AI models, the Model Openness Framework proposes a three-tiered classification for evaluating openness and completeness based on open science principles. Specifically, it breaks down models into constituent code, data, and documentation from a model's development lifecycle and classifies models based on the release of model components under open licenses~\cite{white_model_2024}. Similarly, Garcia et al.\ recommend 10 simple rules for model-sharing based on open science principles~\cite{garcia_ten_2025}. Despite the different approaches of these various frameworks, they share the view that openness in AI R\&D fosters greater transparency, reproducibility, and collaboration.

\subsection{Motivations of open LLM developers}
\noindent  Despite the growth of the open LLM ecosystem, research on motivations remains limited. To date, grassroots communities like EleutherAI and BigScience have articulated motivations rooted in transparency, academic freedom, and public benefit, positioning their work as a means to democratise access to advanced AI capabilities~\cite{akiki_bigscience_2022,ding_towards_2023}. By contrast, Meta has explicitly framed its Llama strategy around ecosystem development, with CEO Mark Zuckerberg explaining: \textit{``We're doing it because... this is an ecosystem... I just want everyone to be using it because the more people who are using it, the more the flywheel will spin for making Llama better''}~\cite{south_park_commons_mark_2024}. The strategy to attract developers to corporate ecosystems was evidenced by the leaked Google memo, in which Google developers warned that they had \textit{``no moat''} against open source and advocated for \textit{``own[ing] the ecosystem''} by \textit{``letting open source work for us''}~\cite{patel_google_2023}. 

However, these observations remain largely anecdotal, and systematic empirical investigation of the motivations driving participation in open LLM development has yet to be conducted. In light of this, prior work on the motivations in OSS development can inform our analysis of open LLM developers' motivations. 

At the individual level, OSS developers are driven by diverse factors, including personal interest, ideological commitments, career development, and business opportunities~\cite{linaaker2024sustaining,von_krogh_carrots_2012}. Importantly, motivations vary significantly based on whether they volunteer or are paid and their geographic context~\cite{lakhani_why_2003,subramanyam_freelibre_2008}. Evidence of variations by geographical regions led Hossain to argue that, \textit{``researchers studying open source should be mindful of geographic variation in what motivates participation and what forms participation may take, particularly outside of the code repository''}~\cite{hossain_regional_2021}. 

Corporate participation in OSS development is typically driven by strategic considerations based on business objectives~\cite{linaaker2020share,li_systematic_2024}. Primary motivations include reducing costs by leveraging community contributions and avoiding duplication of effort~\cite{birkinbine_incorporating_2020,crowston_freelibre_2012}, enabling the growth of business ecosystems around their products and platforms to create network effects and market advantages~\cite{osborne_characterising_2025,srnicek_data_2022}, and influencing the development of open standards that favour their technological approaches~\cite{fink_business_2003,lerner_simple_2002}. Additional strategic benefits include accelerating speed to market~\cite{ahlawat_why_2021,chesbrough_measuring_2023}, reducing dependence on dominant software vendors and avoiding vendor lock-in~\cite{chesbrough_measuring_2023,lerner_simple_2002}, improving corporate reputation within OSS developer communities~\cite{osterloh_trust_2003,osborne_open_2024}, and recruiting talent from OSS developer communities~\cite{agerfalk_outsourcing_2008,fink_business_2003}.

Academic and scientific OSS developers operate within different institutional contexts that shape their motivational frameworks. Academic developers are typically motivated by commitments to transparency, reproducibility, and open science principles that align with scholarly values and institutional expectations~\cite{hasselbring2020open}. Their participation often serves research dissemination goals, enables collaborative scientific advancement, and fulfils growing institutional mandates for open research practices. Public sector motivations centre on the principle that publicly funded software should be publicly available, alongside broader goals of ensuring transparency, maintaining digital sovereignty, spurring innovation, promoting economic growth, and increasing vendor competition~\cite{blind_impact_2021,osborne_european_2023}. Increasingly, governments are also funding OSS development as a strategic investment to enhance national competitiveness in emerging technologies~\cite{osborne_public-private_2024}.

While the OSS literature provides a theoretical foundation for understanding motivations in open LLM development, we note that important distinctions between OSS and open LLM development (e.g., specialised technical expertise, resource requirements, multitude of practitioner communities and artefacts, risks, etc) warrant dedicated investigation.

\subsection{Governance and coordination in open LLM projects}
\noindent  Governance in open LLM projects concerns how authority to shape the project and make decisions is established and exercised within project communities. This includes decision-making around the LLM itself and related artefacts such as data, software, and who can join the collaboration and what roles they are able to play. Given the limited research on governance in open LLM projects, we can consider frameworks from the OSS governance literature, which provide guidance on how collaborative projects coordinate development, make decisions, and manage communities. 

OSS governance structures vary primarily based on how authority is distributed within projects, ranging from autocratic models with centralised leadership to democratic approaches driven by community consensus~\cite{de2007governance,de2013evolution}. Projects can be characterised along a spectrum from community-driven (where the community collectively owns and governs the project) to commercially-driven (where a single entity maintains control while enabling external contributions)~\cite{capra2008framework}. Common governance approaches include meritocratic systems, where influence increases with the quality and quantity of contributions~\cite{butler_investigation_2018}, and foundation-based governance structures that provide neutral oversight for larger projects that require coordination across multiple stakeholders~\cite{mahoney2005non}. These models have evolved to address challenges around resource allocation, technical decision-making, community management, and long-term sustainability that are common across collaborative OSS development efforts.

However, open LLM projects face additional governance challenges that warrant consideration:

\begin{itemize}
    \item \textit{Strategic factors} include defining project scope, vision, and intended applications—whether targeting general-purpose use or specific domains (healthcare, finance), and whether prioritising commercial integration, scientific research, or public access~\cite{widder_open_2023,lawson2023open}.
    \item \textit{Organisational factors} regard the distribution of governance, e.g., to a single organisation (e.g., Meta and their Llama model), to a neutral and collectively owned steward (the RWKV project hosted by the Linux AI \& Data foundation), or informally and decentralised to all actors within the broader ecosystem (e.g., the BigScience Workshop and its BLOOM model) (cf.~\cite{runeson2021open}). 
    \item \textit{Cultural factors} regard the will and potential for fostering an open collaboration within the ecosystem~\cite{heltweg_systematic_2023} and how to structure collaboration for multi-cultural and multi-disciplinary teams~\cite{akiki_bigscience_2022}. 
    \item \textit{Legal factors} encompass the application of different legal regimes of the different components of an AI model or system~\cite{white_model_2024}, managing copyright conditions of the dataset used for training and model outputs~\cite{longpre2023data}, along with legislative requirements on the preservation of privacy (e.g., GDPR in the EU) and transparency and governance~\cite{wagner2023navigating}. 
    \item \textit{Ethical factors} consider the implications and potential use cases of an AI model. For example, while some ecosystems may reside on trust and community norms~\cite{phang2022eleutherai}, the use of responsible licensing, such as \href{https://huggingface.co/blog/open_rail}{OpenRAIL}~\cite{akiki_bigscience_2022} , others have shown how beliefs in technology neutrality trump the potential use cases that the technology may enable~\cite{widder2022limits}. 
    \item \textit{Business factors} concerns the need for business models and innovation to promote co-opetition while enabling sustainable development and provisioning of services based on the open source AI system~\cite{shrestha2023building}, while abiding with restrictions and calls for transparency~\cite{widder_open_2023}.
\end{itemize}




\subsection{Summary}
\noindent  Overall, this review of prior work reveals that critical gaps remain in our understanding of open collaboration methods throughout the lifecycle of open LLMs. First, collaboration practices across the full LLM lifecycle---from initial development through to reuse---have not been systematically mapped (RQ1). Second, evidence of individuals' and organisations' motivations remains mostly anecdotal, which warrants a systematic, empirical analysis (RQ2). Third, the governance and coordination mechanisms that enable successful open LLM collaborations remain unstudied (RQ3). This study addresses these gaps through an exploratory analysis of open collaboration in 14 open LLM projects.

\section{Methods \& Data}\label{sec:methods}
\subsection{Research aims and scope}

\noindent  The goal of this study is to contribute to a more comprehensive understanding of how open collaboration is practised and how it can be further enabled throughout the open LLM lifecycle, guided by RQ1-3 mentioned in the Introduction. 

We concentrate on LLMs given their importance in the open source AI ecosystem since 2021, which provides a larger sample size and longer timeline of activity available for study compared to other types of models, such as multi-modal models. Furthermore, we concentrate on LLMs that have been shared on Hugging Face Hub, which provides relevant information about model release and reuse activity. To understand different practices for models that fall across the spectrum of model openness and licensing, our study includes open LLMs released under permissive licenses (e.g., Apache 2.0 or MIT) and open-weight LLMs released under restrictive licenses (e.g., research-only, acceptable use restrictions or commercial terms). For convenience, we refer to these models as open LLMs in this study.\footnote{N.B. We use the abbreviation ``open LLMs'' as a catch-all term for the 14 LLMs in our sample for convenience only. We explicitly do not intend to make any normative claims about the definition of an ``open LLM.''}


\subsection{Research design}
\noindent  We adopt a qualitative research approach to understand the socio-technical dynamics of open collaboration in the context of LLM development and reuse. Our qualitative research design is suitable for gathering anecdotal knowledge from individuals situated in the problem context, while not necessarily limiting us to a single or selective few cases. Furthermore, we anticipated that practices, norms and culture of collaboration would differ across LLM projects, organisational contexts, and geographies, and therefore qualitative interviews enabled us to capture this diversity.

\begin{sidewaystable*}
\centering
\small
\caption{Summary information about sampled open LLM projects}
\label{tab:interviewees}
\begin{tabular}{>{\centering}p{0.4cm}>{\centering}p{2.41cm}>{\centering}p{2.4cm}>{\centering}p{2.4cm}>{\centering}p{2.4cm}>{\centering}p{2.4cm}>{\centering}p{2.4cm}>{\centering}p{2.4cm}>{\centering\arraybackslash}p{2.4cm}}\toprule
\textbf{ID} & \textbf{Organisation} & \textbf{Organisation Type} & \textbf{Base Models} & \textbf{Derivative Models} & \textbf{Open Source \newline Software} & \textbf{Open Datasets} & \textbf{Other artefacts} & \textbf{Community Platforms} \\
\midrule
I1-I2 & AI Singapore & Public research institute & SEA-LION v1, Llama 3-3.1, Gemma 2-3 & SEA-LION v2-4 & & 18 datasets & Research papers, leaderboard, website, playground, data hub & HF, GitHub, Ollama, Kaggle, Discord, Reddit, LI, Telegram, X \\
\rowcolor{gray!15}
I3 & EleutherAI & Grassroots initiative & Pythia & & LM evaluation harness, GPT-NeoX & 226 datasets & Research papers, blog, evaluations, software & Discord, GitHub, X \\
I4 & BigScience Workshop & Grassroots initiative & BLOOM & & & 10 datasets (e.g. ROOTS) & Research papers, blog & Slack, GitHub, Notion, GDrive \\
\rowcolor{gray!15}
I5 & Cohere Labs & Non-profit arm of startup & Command & Aya-23, Aya Expanse & & 18 datasets & Research papers, blog & Discord, mailing list \\
I6 & Meta & Corporation & Llama v1-4 & & & 11 datasets & Research papers, blog, cookbooks, videos  & GitHub, Hugging Face \\
\rowcolor{gray!15}
I7 & Beijing Academy of AI (BAAI) & Non-profit research institute & Aquila, Emu, OpenSeek, BGE & OPI-Llama-3.8B-Instruct & FlagOpen ecosystem & 111 datasets (e.g. CCI data) & Research papers, leaderboard & HF, X, Modelscope, Github, Gitee \\
I8 & Allen Institute for AI (Ai2) & Non-profit research institute & OLMo-2 & olmOCR, OLMo-2 variants, OLMoE & OLMES, Dolma Toolkit, OLMo-Core, Open Instruct & 265 datasets & Research papers, blog, evaluations & HF, GitHub, Discord, X, LI, Reddit \\
\rowcolor{gray!15}
I9 & Hugging Face & Startup & SmolLM & SmolLM-VS & nanotron, transformers, TRL, lighteval & 49 datasets (e.g. FineWeb 2) & Research papers, blog & HF, Discord, X, LI \\
I10 & SpeakLeash Foundation & Non-profit organisation & Mistral 7B & Bielik V1-3 & ALLaMo & 1 dataset & Blog, benchmarks, Open PL LLM Leaderboard, Chat Arena, Obywatel Bielik app & GitHub, HF, X \\
\rowcolor{gray!15}
I11-I13 & Typhoon \newline (SCB 10X) & VC \& innovation arm of SCB bank & Mistral-7B, Llama 3/Qwen 1.5, Llama 3/Qwen 2, Gemma 3 & Typhoon v1-2 & & 26 datasets (e.g. thai exams data) & Research papers, blog, demo apps & HF, GitHub, X, Discord \\
I14 & Ant Group & Corporation & Qwen3 & inclusionAI, AReal (Ant Reasoning RL) & & ABench & & HF, GitHub, Discord, X \\
\rowcolor{gray!15}
I15 & AI Lab, National Library of Norway & Public sector body & Whisper, BERT & NB-Whisper, nb-bert-base & & 23 datasets (e.g. NCC Corpus) & Demos & HF, GitHub \\
I16 & Masakhane & Grassroots initiative & Llama-3.1-8B & Lugha Llama & & 25 datasets & Research papers, blog & Slack, mailing list, Google groups, \\
\rowcolor{gray!15}
I17 & OpenGPT-X, \newline Fraunhofer IAIS & Research consortium & Teuken-7B & Teuken-7B-v0.4, Teuken-7B-v0.6 & Modalities & 0 datasets & Research papers, blog & HF, LI \\
\bottomrule
\end{tabular}

\small \textit{N.B. This information was provided by the interview respondents. The open dataset values correspond to N datasets hosted on Hugging Face Hub at the time of the research.}
\end{sidewaystable*}

\subsubsection{Interviewee sampling}
\noindent  In line with our exploratory research aims, we employed purposeful sampling to identify representatives from mature (beyond hobby-level), yet distinct cases of open LLM projects that involved open collaborative development during their development and/or reuse lifecycle~\cite{patton2014qualitative}. To create a starting set for our sample, we initially extracted LLMs from the Hugging Face API by filtering for ``\textit{text-generation}'' models and only parent base models that had been finetuned, quantised, or adapted more than 100 times. These variables indicate a maturity and activity in the base model's development and reuse, and potentially also in the child models. Then, we filtered the dataset down to the 250 most downloaded models. We chose the number of downloads metric from Hugging Face API as a proxy for the activity and popularity of a model. The number of downloads is considered for the last 30 days, historically from the day the API is queried. Other metrics, such as likes, are cumulative and may give a skewed distribution toward models that have been on the platform for a longer period of time, but not necessarily actively used today. 

The initial sampling yielded 250 models with a rich set of metadata obtained from the Hugging Face Hub API (see supplementary material). All models represented a finetuned child model to a parent base model. The models were categorised based on the entity that released the model to the Hugging Face Hub, and each author analysed a subset of the models to characterise the type of releasing entity. The analysis leveraged online sources that could be connected with each entity, such as Hugging Face Hub, GitHub, the entities' websites, news media, and social media postings. The categories were formulated through group discussions, and each categorisation was cross-confirmed by a second author. Any disagreements were settled through the joint discussions. The categories that emerged were:

\begin{itemize}
 \item \textbf{Large enterprises}, including corporate organisations with a staff count equal to or more than 250.\footnote{\url{https://single-market-economy.ec.europa.eu/smes/sme-fundamentals/sme-definition_en}}
 \item \textbf{Small- and medium-sized enterprises}, including organisations with a staff count less than 250.
 \item \textbf{Research institutes}, including government-owned or private universities and research institutions.
 \item \textbf{Non-profit / grassroots entities}, including research-focused organisations without business models and revenue streams connected to their research.
 \item \textbf{Individuals}, including individuals who have shared the model in their private capacity.\end{itemize}

To filter for LLMs with the potentially highest levels of activity and socio-technical collaboration, we leverage the number of downloads as a proxy for technical activity and the number of discussions in the model's Hugging Face Hub repository as a proxy for social activity per model. These are rough indicators but serve as an initial proxy to attain an ordered list for the continued sampling. Upon considering the diversity of organisation types in our sample and verifying the willingness of representatives to be interviewed, we included the LLMs represented by interviewees I1-2 and I5-8 in our sample (see Table~\ref{tab:interviewees}). However, something that stood out in this initial dataset and sample was the absence of community-driven open LLM projects, which our team was aware of based on our contextual awareness of the open source AI ecosystem. In line with our purposeful sampling approach, we complemented the initial sample with the LLMs represented by I3-4 and I9-17 for a fuller picture of the open LLM landscape. Further information about the sampled models and the organisations driving their development can be found in~\ref{app:cases}.

\subsubsection{Interview process}
\noindent  Based on these insights from our analysis of the initial sample, we constructed an interview guide comprising questions about collaboration practices, on-ramps, and challenges throughout the open LLM lifecycle (RQ1); motivations (RQ2); and how open LLM collaborations are coordinated and governed (RQ3).
Subsequently, we recruited interviewees for the target open LLM projects via email outreach and personal networks. Interviewees were asked to confirm their informed consent before participating in the interview. The interviews were primarily conducted online with one or two of the authors present, who took turns in asking questions and taking notes. A select number of interviews (I2, I5, I9) were conducted in person, following a similar format. The interviews were semi-structured, combining questions from the interview guide and spontaneous queries, and typically lasted 30-60 minutes. The interviews were recorded for transcription, and the recordings were transcribed using offline transcription software and managed in accordance with a data management plan. 


\subsubsection{Data analysis}
\noindent  After each interview, transcripts were coded by one of the authors using an abductive coding approach. The code book 
was initially constructed in alignment with the interview guide and the LLM lifecycle framework~\cite{ding_2024_pipeline}. After an interview transcript had been coded, a second author reviewed the codes, after which these and any other reflections were discussed and shared among the author team. After the third interview, the first-level codes began to be categorised into second-level codes (or themes), which were iteratively refined and revised following joint discussions. The code book includes a general section of themes touching on the interviewees’ overarching understanding of the open LLM ecosystem, incentives and drivers, and experience of open collaborative development of their open LLM. 

The dual coding approach was iterated for all interviews to mitigate researcher bias. We also conducted peer debriefings at regular intervals to share our findings and reach a collective understanding. After I10, we summarised our findings in a tentative synthesis. While we agreed that the findings were starting to saturate — e.g., as no additional high-level themes had emerged in recent interviews — we agreed to extend our investigation with I11-17 to broaden the diversity of our sample. After coding the 17th interview, we agreed that saturation had been reached and concluded data collection and analysis.



\section{Collaboration Challenges and On-ramps throughout the Open LLM Lifecycle (RQ1)}\label{sec:results-collabs}

\begin{sidewaysfigure*}
    \centering
    \includegraphics[width=1\linewidth]{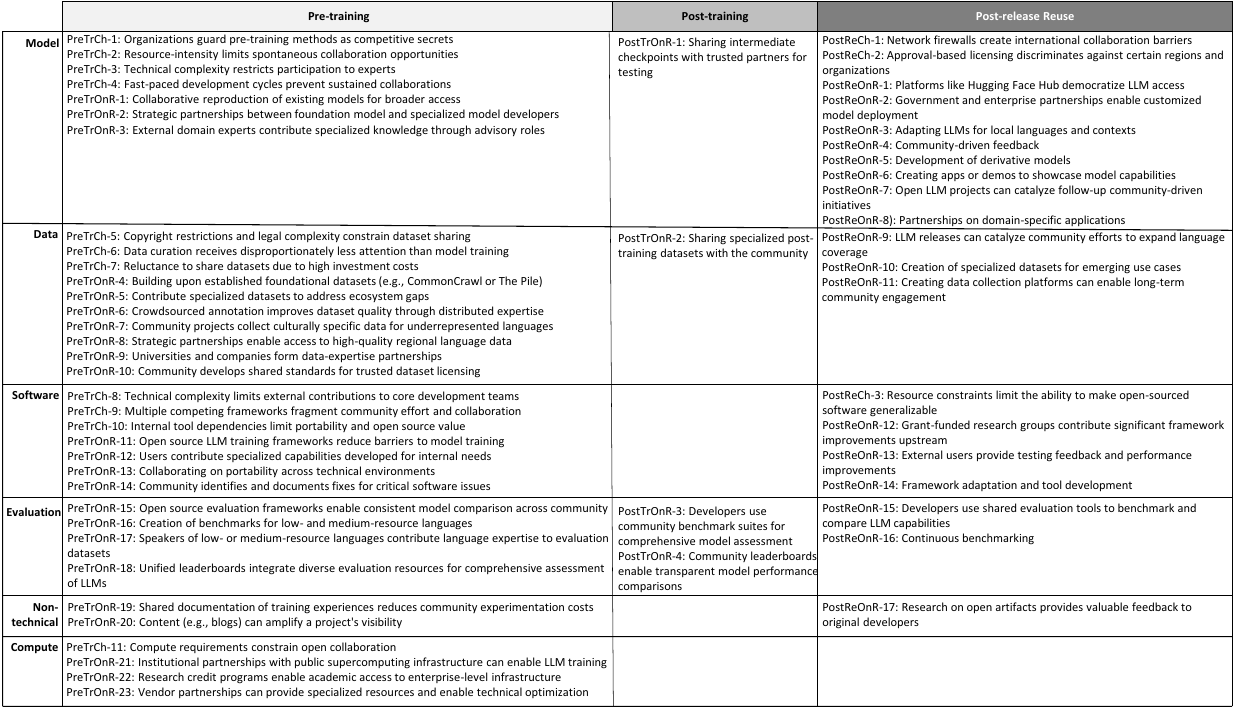}
    \caption{Overview of collaboration Challenges (Ch) and Cn-ramps (OnR) per phase (pre-training (PreTr), Post-training (PostTr), Post-release Reuse (PostRe)) and area of collaboration (model, data, software, evaluation, non-technical, compute). Challenges and On-ramps are abbreviated [Phase][Challenge/On-ramp number in the phase].}
    \label{fig:framework}
\end{sidewaysfigure*}

\noindent  In this section, we present our findings on where and how open collaboration occurs at each stage of the open LLM lifecycle, highlighting collaboration on-ramps and challenges at each stage. The Challenges (Ch) and Cn-ramps (OnR) are synthesised per phase (pre-training (PreTr), Post-training (PostTr), Post-release Reuse (PostRe)) and area of collaboration (model, data, software, evaluation, non-technical, compute). Challenges and On-ramps are abbreviated [Phase][Challenge/On-ramp number in the phase]. An overview is provided in Fig.~\ref{fig:framework}, and a more thorough description and reporting of each challenge and on-ramp is provided in~\ref{app:collab-challenges-and-onramps}.

\subsection{Pre-training Stage}

\begin{framed}
\noindent $\diamond$ \textbf{Summary:}
   \noindent Open collaboration in the pre-training stage of the LLM lifecycle encompasses a variety of artefacts and activities that lead up to the generation of a pre-trained (base) model, spanning the curation of datasets, contributions to open source training frameworks (e.g., a model architecture), non-technical tasks (e.g., documentation, community outreach), and forming partnerships for compute access.  \newline 
    
    \noindent $\diamond$ \textbf{Collaboration challenges:} During the pre-training stage, open collaboration is often limited by proprietary considerations, regulatory and geographical constraints, and the complexity of planning shared activities in the resource- and expertise-intensive process of pre-training an LLM. In some cases, model training processes are treated as corporate secrets, which restricts the possibility of knowledge sharing and open collaboration. In the context of training datasets, copyright restrictions and a structural undervaluation of data curation work further impede collaborative efforts. In addition, there is often a limited appetite to share curated training datasets due to their high costs and perceived value. Software tools and infrastructure for pre-training models are commonly tailored to internal systems, and the technical, organisation-specific expertise required for contributions limits the scope of external involvement. Lastly, open LLM developers, particularly grassroots projects, struggle to access the computing resources needed for pre-training LLMs. \newline
    
    \noindent $\diamond$ \textbf{Collaboration on-ramps:} We identify several on-ramps that facilitate collaboration during the pre-training stage. Opportunities include reproducing existing models, finetuning models for specific use cases, and integrating model architectures into open source LLM training frameworks. Data collaborations can be formed when there are shared interests in producing a resource useful to many. This takes the form of reusing existing datasets, releasing internally developed datasets, developing data curation partnerships, and engaging in community-driven data collection, e.g., for low- or medium-resource languages. There is also collaboration to develop open source frameworks for training and evaluating LLMs. Non-technical contributions — such as sharing knowledge, model cards, and best practices — support learning and facilitate the advancement of research ideas and practices. Additionally, developers form partnerships (e.g., with public and corporate providers) to secure access to the compute resources needed to train open LLMs.  
\end{framed}

\subsubsection{Pre-training: Model Collaborations}
\noindent  Formal partnerships between and across communities and organisations provide essential enablers and mediators for model collaborations. For example, developers of base and derivative models collaborate formally and informally to enhance both model capabilities and regional representation (\textbf{PreTrOnR-2}). These collaborations involve information sharing, early access arrangements, and data contributions and partnerships that benefit the broader ecosystem. AI Singapore has engaged with both the Llama team at Meta and the Gemma team at Google, with different levels of collaboration in each case. Another example of collaboration regards the reproduction of existing models to enable broader access, e.g., due to licensing restrictions (\textbf{PreTrOnR-1}). These reproduction efforts leverage shared codebases and architectural similarities to accelerate development. For example, EleutherAI and Hugging Face collaborated on implementing Meta's first Llama model.

Some organisations, however, tend to treat information about their model development processes as a competitive advantage, creating barriers for collaboration (\textbf{PreTrCh-1}). The resource-intensive nature of LLM pre-training further requires extensive upfront planning that can stifle organic collaboration opportunities (\textbf{PreTrCh-2}). Developers must scope potential collaborations early in the development cycle because late-stage partnerships can fundamentally alter experimental approaches and resource allocation. Additionally, the fast-paced nature of open LLM development makes it difficult to maintain long-term engagement and build lasting collaborative relationships in open LLM projects (\textbf{PreTrCh-4}).

Another critical barrier to model collaboration is the technical complexity, which restricts participation to experts (\textbf{PreTrCh-3}). This expertise bottleneck is particularly pronounced in regions with fewer AI specialists. I10 from the SpeakLeash Foundation highlights this challenge: \textit{``the training is [done] mostly by our internal team. Because one thing is that this is very specialised knowledge and not everyone has it... And currently, I believe we do not have many more experts on this level in Poland.''} One approach suggested for addressing the limited access to expertise is to engage external domain experts to contribute specialised knowledge through advisory roles (\textbf{PreTrOnR-3}). This enables projects to access high-level guidance without requiring contributors to make a full-time commitment. For example, I16 from Masakhane describes how, \textit{``[they] asked for technical advice... and one professor from the U.S. was interested [and had] been doing machine translation for many years.''}

\subsubsection{Pre-training: Data Collaborations}
\noindent  Data collaborations during pre-training encompass dataset creation, curation, and validation activities that are essential for model development but pose distinct challenges compared to model collaborations. Copyright restrictions and rights-holder interventions create significant legal complexity that constrain both dataset creation and model release strategies, leading developers to implement careful filtering and dual-licensing approaches to ensure compliance with the law (\textbf{PreTrCh-5}). This is further complicated by the fact that data preparation and curation receive disproportionately less attention and resources than model training, creating a systemic imbalance that affects dataset quality and availability (\textbf{PreTrCh-6}). Interviews, however, highlight that developers can and do collaborate to address data provenance and mis-licensing issues, for example, by developing shared standards and validation processes that build trust in open training datasets (\textbf{PreTrOnR-10}). Large-scale community engagement that enables organisations to improve dataset quality through distributed annotation and feedback has also been proven possible, particularly valuable for multilingual datasets where native speaker expertise is essential (\textbf{PreTrOnR-6}). Addressing data gaps for underrepresented languages is iterated across interviews as an opportunity for collaboration (\textbf{PreTrOnR-7}), while requiring careful coordination to establish shared quality standards and cultural guidelines across diverse contributor communities.

Despite high demand for quality datasets, organisations remain reluctant to share curated data due to the substantial time investment required to prepare them and the perceived competitive value, creating an asymmetric ecosystem where data consumption exceeds contribution (\textbf{PreTrCh-7}). Strategic partnerships appear as key for inter-organisational sharing, where conditions are defined up-front (PreTrOnR-8). These partnerships often extend beyond simple data exchange to include technical collaboration and capacity building. For example, AI Singapore collaborates with Google Research to \textit{``collecting and transcribing voice audio specifically for Southeast Asian languages,''} while also working with regional partners to encourage data release by emphasising social-good incentives. Collaborations between companies and academic labs are specifically highlighted as an opportunity for mutually beneficial arrangements in which companies provide data resources and academic labs contribute research expertise and validation (\textbf{PreTrOnR-9}).

A means of addressing limited sharing is to build on established open datasets like CommonCrawl and The Pile, which serve as foundational sources that enable collaboration across the ecosystem (\textbf{PreTrOnR-4}). These data sources require different levels of processing, with some providing pre-cleaned data while others need extensive filtering and preparation before they can be used to pre-train a model. I7 from BAAI explains their approach: they acquire training data from \textit{``two primary sources: CommonCrawl and the Pile datasets,''} noting that \textit{``CommonCrawl datasets require extra processing, while the Pile datasets are already pre-cleaned.''} Creating or recreating specialised datasets not necessarily of competitive advantage is also highlighted as a means of addressing gaps in existing training data in the ecosystem, particularly for specific capabilities or domains that lack adequate coverage (\textbf{PreTrOnR-5}).

\subsubsection{Pre-training: Software Collaborations}
\noindent  Software collaborations in pre-training focus on developing LLMs, in particular, training and evaluation frameworks. Developers engineer open source training frameworks to enable broader community access to LLM training capabilities while benefiting from community feedback and contributions (\textbf{PreTrOnR-11}). These frameworks become foundational infrastructure that reduces barriers for other teams to train models. As these frameworks are commonly considered commodities, developers contribute features to them when developing capabilities for their own use cases (\textbf{PreTrOnR-12}). The area is still maturing, however, as there is an absence of standardised frameworks for training and evaluating LLMs, which results in ecosystem fragmentation and duplication of effort across the community (\textbf{PreTrCh-9}). While the diversity of frameworks can serve different use cases, the lack of convergence fragments collaboration. Still, interviewees report that much infrastructure is developed internally and not shared due to internal tool dependencies, which limit portability and open source value (\textbf{PreTrCh-10}).

On the other hand, the opportunity to collaborate on the portability of frameworks across technical environments is highlighted as a prime opportunity. The AI lab at the National Library of Norway has collaborated with Hugging Face to port PyTorch code to JAX to meet their specific technical requirements. I15 explains, \textit{``we did the port of their code to JAX. They had it in PyTorch, and we really needed that code in JAX... That was a lot back and forth working directly on the code with feedback on the exact part of the code and how to do things.''} While their internal code contained dataset-specific features that were unsuitable for inclusion in the general codebase, the core porting work benefited both organisations and the broader community.

A general challenge limiting contributions is the technical expertise and effort required to develop and maintain LLM training tools and infrastructure, with most development remaining centralised within core teams (\textbf{PreTrCh-8}). I3 from EleutherAI explains that while EleutherAI's GPT NeoX library has received some external contributions from researchers and developers, \textit{``who are using it professionally, either in academia or in industry,''} such contributions are rare due to expertise requirements. An opportunity for contributions with a lower barrier of entry concerns information-based collaborations in which organisations report and document fixes for bugs identified in core open source software used for model pre-training (\textbf{PreTrOnR-14}). I8 from Ai2 describes a case where their team \textit{``found a bug in how numbers have shuffled in PyTorch,''} and reported it to the team.

\subsubsection{Pre-training: Evaluation Collaborations}
\noindent  Evaluation collaborations during pre-training focus on developing benchmark datasets, frameworks, and tools that serve the community's needs for consistent model capability comparison. The development of evaluation frameworks is highlighted as an important area of collaboration, serving shared community needs for consistent, reproducible model assessment while providing platforms for researchers to contribute new benchmarks and evaluation protocols (\textbf{PreTrOnR-15}). EleutherAI's LM Evaluation Harness is an example of this model, enabling comparison of models across evaluation datasets in a \textit{``consistent, reproducible, and reliable''} manner. Evaluation dataset development also represents a major collaborative opportunity that brings together diverse partners, including academic institutions, language communities, and model developers, to create comprehensive benchmarking frameworks for specific linguistic or cultural contexts (\textbf{PreTrOnR-16}). Language communities and domain experts contribute to the creation of evaluation datasets through crowdsourced efforts that capture cultural and linguistic nuances essential for comprehensive model assessment (\textbf{PreTrOnR-17}). These collaborations leverage community knowledge to identify and digitise evaluation materials that would otherwise remain inaccessible. Another important area for collaboration raised includes unified leaderboards that integrate diverse evaluation resources for comprehensive assessment of LLMs (\textbf{PreTrOnR-18}).

\subsubsection{Pre-training: Non-technical Collaborations}

\noindent  Non-technical collaborations during pre-training focus on knowledge sharing, community engagement, and research dissemination, supporting the broader ecosystem beyond code and data contributions. Developers contribute to the community's collective knowledge base by sharing experimental methodologies, evaluation practices, and detailed training logs that enable others to learn from their experiences and avoid common pitfalls (\textbf{PreTrOnR-19}). This knowledge sharing is particularly valuable given the complexity and resource intensity of LLM pre-training, where learning from others' approaches can significantly reduce experimentation costs. I8 from Ai2 emphasises the importance of such sharing: \textit{``Specifically, when it comes to building, the more compute you want to put in your effort, the more design space what you can do to achieve your goal is like huge, right? And so learning how others either have approached similar problem or start or their final end stage of starting point, or just in general how they think about problem. It is very valuable.''} Beyond documentation contributions, general content production and community outreach efforts also provide an important area for collaboration (\textbf{PreTrOnR-20}). These types of contributions help promote awareness of technical contributions, share success stories, and build engagement around collaborative projects. This includes coordinated content creation and marketing efforts that amplify the impact of technical collaborations.

\subsubsection{Pre-training: Compute Collaborations}

\noindent  Compute access represents one of the most critical barriers and collaborative opportunities in open LLM development, given the substantial computational resources required for training LLMs (\textbf{PreTrCh-11}). Model development becomes limited to organisations with substantial internal resources or partnership arrangements. This resource constraint affects even well-funded organisations and extends beyond raw compute to include the financial and technical complexity of managing large-scale training infrastructure. I3 from EleutherAI emphasises the universal nature of this challenge: \textit{``Resource access is a really big deal. Most organisations do not have on hand the computing resources required to train one of these models. So, figuring out how to make that happen is really important.''} Fortunately, interviews demonstrate that national and institutional supercomputing centres actively seek partnerships with open LLM developers to validate their infrastructure while providing essential compute resources for large-scale model training projects (\textbf{PreTrOnR-21}). These partnerships often emerge from mutual benefits: computing centres need stress-testing and benchmarking, while developers need access to cutting-edge hardware. The BigScience Workshop, for example, was enabled to train its BLOOM model through an invitation from Jean Zay, a French public cluster, while the SpeakLeash Foundation was enabled to train its Bielik models through Poland's Cyfronet supercomputing centre.

An opportunity for cloud providers is to offer research credits and specialised programs that enable academic and open source projects to access enterprise-level compute infrastructure, often in exchange for open research publication requirements (\textbf{PreTrOnR-22}). Beyond credits, open LLM developers have formed partnerships with hardware and cloud providers to access compute for training LLMs. These partnerships have involved technical collaboration on setup and optimisation in addition to the provision of compute resources (\textbf{PreTrOnR-23}).

\subsection{Post-training Stage}

\begin{framed}
\noindent $\diamond$ 
\textbf{Summary: }
\noindent We observe relatively little open collaboration during the post-training stage of open LLMs, as most post-training activities remain internal to organisations. However, the collaborations that do occur include sharing intermediate model checkpoints for testing and feedback with trusted partners, releasing curated post-training datasets, and leveraging open source benchmarks for model evaluation pre-release. \newline

\noindent $\diamond$ \textbf{Collaboration challenges:} The post-training stage involves significant barriers to open collaboration. The technical complexity and resource intensity of post-training methods create high barriers to entry for external collaborators, while the competitive sensitivity around model capabilities and performance optimisations limits willingness to share intermediate training states or detailed methodologies. In addition, the rapid iteration cycles common in post-training make it challenging to coordinate external input, and the lack of standardised frameworks for collaborative post-training further constrains community involvement. \newline

\noindent $\diamond$ \textbf{Collaboration on-ramps:} Despite limited collaboration during post-training, we observe at least three on-ramps for collaboration. First, the sharing of intermediate checkpoints with trusted partners for testing and feedback. Second, the curation of specialised post-training datasets. Third, the curation of evaluation datasets and the use of community-managed evaluation resources (e.g., open benchmarks, LLM leaderboards).

\end{framed}

\subsubsection{Post-training: Model Collaborations}
\noindent  Post-training model collaborations are currently relatively limited, focusing on performance testing and feedback collection. Interviewees report on how developers may share intermediate model checkpoints with selected partners to gather specialised feedback and test specific capabilities during the post-training process, enabling iterative improvement before the public release of the model (\textbf{PostTrOnR-1}). For example, I9 from Hugging Face explains that they had \textit{``shared the intermediate checkpoint of the model with other startups, so that they could test them and then see if we could add new capabilities to the model during post-training.''} Through these targeted collaborations, they \textit{``got some feedback about the models and then had the final instruct model released,''} demonstrating how selective sharing enables quality improvement through external validation.

\subsubsection{Post-training: Data Collaborations}
\noindent In terms of data collaborations, developers share specialised post-training datasets that enable other developers to improve their models. These releases can also have multiplier effects across the ecosystem (\textbf{PostTrOnR-2}). For example, I7 from BAAI mentions that they had \textit{``released a post-training dataset called Infinity Instruct, which they had developed for post-training their models like Aquila, on Hugging Face Hub. After three months, it had been used by external developers to post-train over 130 models.''I7 explains that her team is satisfied with the collaboration this facilitates, stating that, \textit{We open source our dataset, and a lot of people use that to change their model, and they again open source their model.} So, we are happy to see this kind of open source cycle.''} 

\subsubsection{Post-training: Evaluation Collaborations}
\noindent  Evaluation is the most collaborative aspect of the LLM post-training stage, as developers rely heavily on community-developed benchmarks and evaluation infrastructure to assess model performance and communicate capabilities to users (\textbf{PostTrOnR-3}). This reliance on community resources reflects both practical constraints and best practice recommendations from leading research groups. For example, I9 from Hugging Face describes testing SmolLM \textit{on a very comprehensive list of benchmarks that test really different abilities and also some benchmarks that we did not monitor during the training to make sure ...that we do not overfit on them,''} following practices \textit{recommended by Ai2 in their OLMO paper.''} I7 from BAAI emphasises the necessity of this approach: \textit{if we want to have a full spectrum benchmark for the model,''} it is important to develop benchmarks for various capabilities, but \textit{it is impossible for us to develop all the benchmarks by ourselves.''} Consequently, they \textit{``use a lot of open source benchmarks as part of our evaluation method.''} Community-managed evaluation platforms and leaderboards are another important area for indirect collaboration, where organisations contribute evaluation results while benefiting from shared evaluation infrastructure (\textbf{PostTrOnR-4}).

\subsection{Post-release Reuse Stage}

\begin{framed}
\noindent $\diamond$ 
\textbf{Summary: }
\noindent Open collaboration in the reuse stage encompasses the downstream activities that occur after initial LLM release, spanning dissemination and adoption, derivative development, community feedback mechanisms, and collaborative improvements to associated datasets, software tools, and evaluation frameworks. These post-release collaborations often demonstrate greater openness and community engagement compared to the pre-training stage, with platforms like Hugging Face facilitating widespread distribution and reuse. \newline

\noindent $\diamond$ \textbf{Collaboration challenges:} Model reuse collaborations face several structural limitations, including limited direct engagement between original developers and derivative model creators, with most collaboration happening indirectly through platform-mediated adoption rather than active co-development. Many organisations release code and models but lack the resources to make them truly accessible to the broader community, requiring significant additional effort for documentation, generalisation, and maintenance beyond their primary research goals. Community engagement can be difficult to sustain over time, particularly for volunteer-based initiatives, and developers often struggle to balance supporting diverse use cases while maintaining focus on their core objectives. Additionally, while feedback mechanisms are valuable, they tend to be reactive rather than proactive, limiting opportunities for deeper collaborative development. \newline

\noindent $\diamond$ \textbf{Collaboration on-ramps:} Despite these challenges, the model reuse stage offers numerous pathways for meaningful collaboration, particularly through established platforms that facilitate model dissemination, adoption, and community feedback. Organisations successfully collaborate by adapting and fine-tuning models for specific domains (e.g., local languages, specialised applications), with many derivative projects building on open models such as Llama, Pythia, and SEA-LION. Software tools and evaluation frameworks receive ongoing community contributions, including performance improvements, new features, and specialised benchmarks, with projects like GPT-NeoX and evaluation platforms benefiting from upstream contributions. Non-technical collaborations thrive through research publications that build on open artefacts, providing valuable insights back to the original developers, while compute-sharing initiatives help accelerate community projects and democratise access to the computational resources needed for model development and evaluation.
\end{framed}

\subsubsection{Post-release: Model Collaborations}

\noindent  Model collaborations during the post-release (or reuse) stage encompass the diverse ways developers and communities interact with and build upon released models, ranging from platform-mediated distribution to direct partnerships for specialised applications. While most collaboration occurs indirectly through adoption rather than active co-development, several pathways have emerged that enable meaningful engagement between model creators and users.

Standardised collaboration platforms, such as Hugging Face, serve as essential intermediaries that enable widespread model distribution and community access, effectively democratizing model availability beyond direct organisational relationships (\textbf{PostReOnR-1})s. Open models, released via these platforms, enable distributed practitioners to develop derivative LLMs without direct involvement by the developers of the base LLM (\textbf{PostReOnR-5}). From finetune to experimentation, these community initiatives often explore directions that core teams do not or cannot pursue due to resource constraints. Feedback mechanisms enable community members to contribute feedback about model performance, limitations, and potential improvements (\textbf{PostReOnR-4}). While feedback tends to be reactive rather than proactive, it provides valuable insights for model refinement. LLM releases further can catalyse new collaborative initiatives that build upon lessons learned and resources developed, creating successive waves of community-driven projects with improved methodologies and focus (\textbf{PostReOnR-7}). These follow-up projects often inherit infrastructure and knowledge while addressing limitations identified in earlier efforts.

Partnerships between model builders and various actors are iterated as a key enabler for different types of collaborations. One example is how government and enterprise partnerships enable customised model deployment, creating collaborative relationships that extend beyond simple model distribution to include technical support and customisation guidance (\textbf{PostReOnR-2}). These partnerships often serve as showcase examples that demonstrate model capabilities and encourage broader adoption. Regional partnerships have focused on adapting base models to local languages and cultural contexts, often involving multiple stakeholders from a geographic region working toward common goals such as language representation (\textbf{PostReOnR-3}). AI Singapore exemplifies this approach through the SEA-LION model ecosystem, where partners across Southeast Asia have fine-tuned the models for local applications. For example, organisations in Thailand (Visitec) and Indonesia (GoTo) have created derivatives, with GoTo developing a customer service assistant integrated into their payments app that operates in Javanese, Sundanese, and Indonesian. 

Some organisations have focused on partnerships in specialised domains where model finetuning requires domain expertise (\textbf{PostReOnR-8}). For example, I6 from Meta explains that they have focused on applying Llama to \textit{``specific use cases like legal applications or medical applications.''} Model developers further collaborate with startups and enterprises to create demo applications that showcase model capabilities and explore potential use cases, benefiting both parties through technical validation and market visibility (\textbf{PostReOnR-6}). These partnerships often involve fine-tuning and specialised applications that highlight the versatility of smaller or specialised models.

Despite the many opportunities, some interviewees report on barriers that effectively limit collaboration. Geographical restrictions create friction for international collaboration in open LLM development, manifesting through both network firewalls and discriminatory licensing practices (\textbf{PostReCh-1}). I14 from Ant Group explains how the firewall in China creates collaboration barriers: \textit{we work internally because there is a firewall. So, we just developed an internal model scope or model states in our internal GitLab, and we upload our updates to GitHub and Hugging Face. We have a regular update, for example, like once a week, and we have some major updates like once every two months or three months.''} Another form of barrier regards how approval-based licensing discriminates against certain regions and organisations. Again, I14 from Ant Group describes experiencing this exclusion when attempting to access Llama models: \textit{``I cannot apply for the [Llama] license because my identity says that I'm in mainland China.''} 

\subsubsection{Post-release: Data Collaborations}

\noindent  Data collaborations during the post-release stage focus on creating datasets to improve pre-trained models. Model releases may, for example, catalyse ongoing community efforts to expand language coverage and improve data representation for underrepresented languages, often extending beyond the original project scope through sustained collaborative initiatives (\textbf{PostReOnR-9}). These efforts leverage the infrastructure and momentum established during initial development while addressing gaps identified through community feedback. Open model releases can also inspire the development of specialised datasets that address specific use cases or methodological needs identified through model deployment and community experimentation (\textbf{PostReOnR-10}). These derivative datasets often solve technical challenges that emerged during practical applications, contributing valuable resources back to the broader ecosystem. Some developers have reportedly also established data collection mechanisms following model releases to sustain ongoing community engagement and enable continuous improvement in language representation and model capabilities (\textbf{PostReOnR-11}). These infrastructure investments create long-term collaborative platforms that extend the impact of initial model releases. For example, the Cohere for AI team built a dedicated website for continuously collecting language data following their Aya model releases, creating a sustainable mechanism for community members to contribute linguistic and cultural knowledge that can inform future model development and ensure continued expansion of language coverage and cultural representation.

\subsubsection{Post-release: Software Collaborations}

\noindent  Software collaborations during the post-release stage focus on maintaining and extending open source frameworks and tools. While some organisations face resource constraints in making their software broadly accessible (\textbf{PostReCh-3}), others still find opportunity and value in releasing an internally developed framework as open source (\textbf{PostReOnR-14}). Academic research groups, for example, contribute significant improvements to open source frameworks by adapting tools in their projects, often upstream optimisations and features that benefit the entire community (\textbf{PostReOnR-12}). These collaborations typically emerge from grant-funded research projects that have both the resources and motivation to contribute back to the ecosystem. There is also an opportunity for contributions in the form of testing and feedback from external researchers and developers (\textbf{PostReOnR-13}). These contributions often focus on improving usability, performance, and expanding applicability to new use cases.

\subsubsection{Post-release: Evaluation Collaborations}

\noindent Evaluation collaborations during the post-release stage focus more on knowledge sharing and establishing common practices than on developer-oriented activities. Developers, for example, actively participate in community-managed benchmarking platforms that provide standardised evaluation frameworks and transparent model comparison capabilities, benefiting from shared evaluation infrastructure while contributing to ecosystem-wide performance tracking (\textbf{PostReOnR-15}). These platforms serve as neutral ground for model comparison and help users make informed decisions about model selection. Continuous benchmarking is another area for collaboration, where evaluation platforms establish mechanisms for ongoing community contributions that expand benchmark coverage, update evaluation results, and incorporate new models and capabilities as the field evolves (\textbf{PostReOnR-16}). 

\subsubsection{Post-release: Non-technical Collaborations}

\noindent  Non-technical collaborations focus on research activities that build on open artefacts to generate new insights, creating feedback loops in which external research contributions inform future model development (\textbf{PostReOnR-17}). I8 from Ai2 describes the impact of external scientific research that uses open models and facilitates open model development. For example, they highlight the ``Fishing for Magikarp'' paper by Land \& Bartolo~\cite{land_fishing_2024}, who found that unique token sequences (``glitch tokens'') lead to unpredictable but repeatable model failures, which can be at least partly traced to undertrained tokens, a hypothesis that is much easier to verify for models whose pretraining corpus is fully released. I8 notes that this research provided concrete insights about how \textit{``vocabulary may impact stability of a run during training...So, it was not like a direct contribution, but it was a finding that builds specifically on top of our artefacts, and then it formed like a whole line of exploration that made our models better.''} This demonstrates how transparency enables research that creates unexpected value for original developers.

\section{Motivations for engaging in the collaborative development of open LLMs (RQ2)}\label{sec:results-motives}

\begin{figure*}
    \centering
    \includegraphics[width=0.7\linewidth]{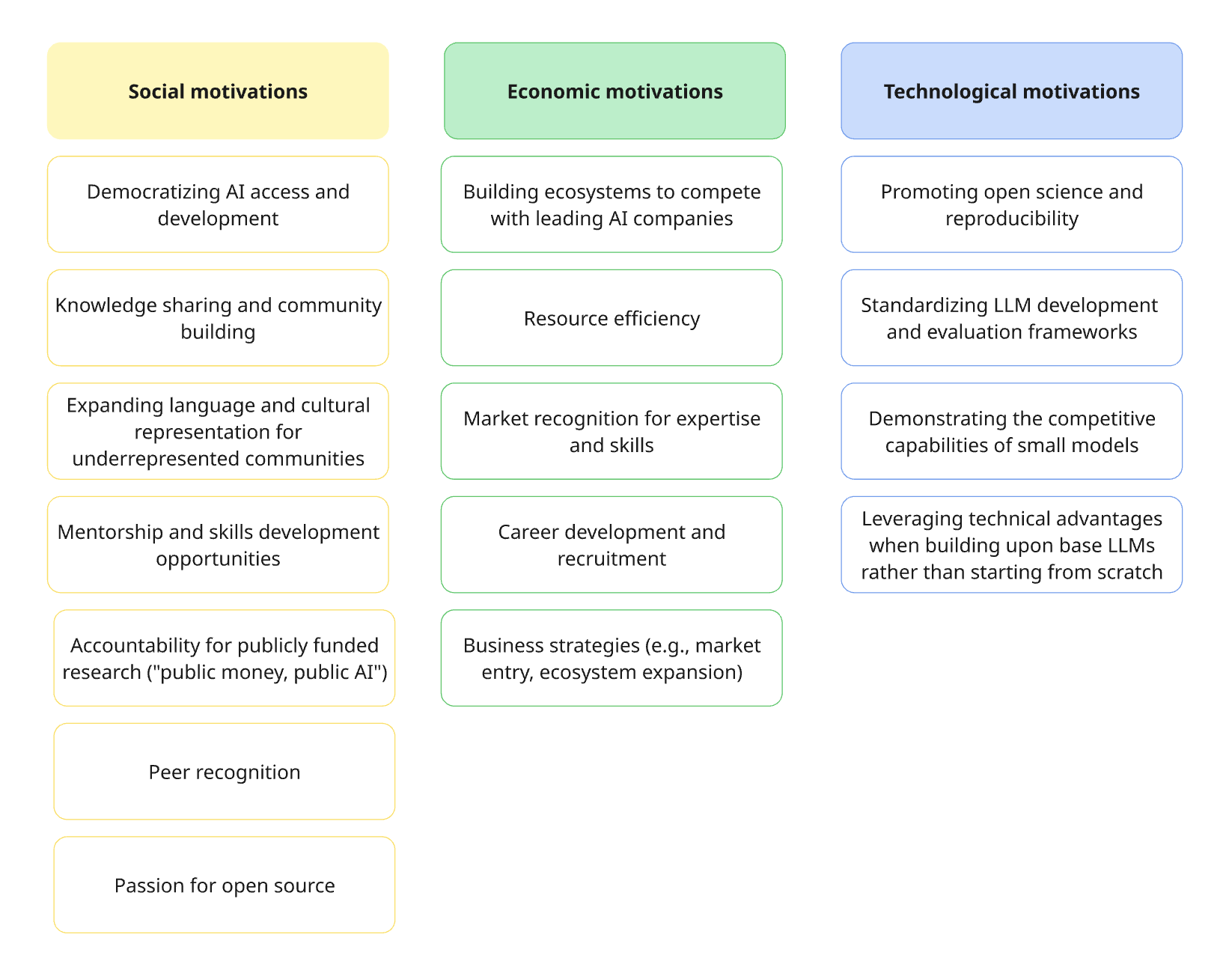}
    \caption{Motivations of Open LLM Developers}
    \label{fig:motivations}
\end{figure*}

\begin{framed}
\noindent $\diamond$ \textbf{Summary:}
\noindent We observe social, economic, and technological motivations that incentivise various stakeholders to participate in the collaborative development of open LLMs and related artefacts, as illustrated in Fig.~\ref{fig:motivations}.\newline 

\noindent $\diamond$ \textbf{Social motivations:} Democratizing AI access and development, knowledge sharing and community building, expanding language and cultural representation for underrepresented communities, providing mentorship and skills development opportunities, ensuring public accountability for publicly funded research (``public money, public AI''), gaining peer recognition, and pursuing one's passion for open source. \newline 

\noindent $\diamond$ \textbf{Economic motivations:} Building ecosystems to compete with leading AI companies, resource efficiency, gaining market recognition for expertise and capabilities, career development and recruitment, and business strategies that seek to enter markets and build ecosystems. \newline 

\noindent $\diamond$ \textbf{Technological motivations:} Promoting open science and reproducibility, standardising LLM development and evaluation frameworks, demonstrating the competitive capabilities of small models, and leveraging technical advantages when building upon base LLMs rather than starting from scratch.
\end{framed}

\subsection{Social Motivations}

\subsubsection{Democratizing access and inclusion in AI development}

\noindent  Democratising access to state-of-the-art AI capabilities and ensuring broad inclusion in AI development are principal motivations for many open source AI developers. Organisations like AI Singapore conceptualise their work as ``public infrastructure'' that should remain open to prevent AI benefits from being concentrated within any single company or country. This democratization approach enables ecosystem building that brings together diverse stakeholders from public entities, philanthropic organisations, and private companies. For example, AI Singapore collaborates with both regional university labs and global technology companies, such as Google, to collect data on Southeast Asian languages. 

Similarly, I4 from the BigScience Workshop emphasises broad inclusion as the project's guiding value, investing significant resources in recruiting contributors beyond their personal networks. Furthermore, I5 from Cohere for AI describes how their community-driven approach brings together researchers from institutions of varying resources and prestige to collaborate, share knowledge, and advance AI research collectively.

\subsubsection{Knowledge sharing and community building}

\noindent  Building a community of practice and enabling knowledge exchange are key motivations for many open LLM projects. This approach facilitates skill sharing, mutual learning, and the creation of more diverse contributor communities, thereby strengthening the overall ecosystem. For example, I4 from the BigScience Workshop highlights how skill sharing and mutual learning were incentives for many volunteers. EleutherAI shares the mission of enabling access to open models while driving research innovation. I3 from EleutherAI describes how they provide platforms for researchers and peers to meet, collaborate, and access necessary resources and infrastructure, with notable experts regularly joining their Discord community to help answer technical questions. 

\subsubsection{Expanding language and cultural representation in LLMs}

\noindent  Expanding language and cultural coverage in LLM pre-training data is another primary motivation for many developers. In particular, these efforts seek to address the historical neglect of medium and low-resource languages in LLM development and ensure that LLMs serve global communities rather than only dominant language groups.

Multiple projects demonstrate this motivation. Cohere for AI's Aya Initiative targeted coverage for 101 languages, while AI Singapore focused on 13 languages and dialects of Southeast Asia. Regional initiatives include SpeakLeash's focus on Polish language and culture, SCB 10X's work on Thai language nuances, and the OpenGPT-X project's development of a multimodal LLM that performs well across all 24 official EU languages. I11 from SCB 10X explains their motivation: \textit{``The main motivation really came from [our head of AI strategy]. He wanted to build a model that was very open and useful to all Thai people. Basically, the way he viewed it is that language is something that is really cultural for us, and having the ability for Thai people to develop their own models, to understand the cultural nuances of Thai and to be able to integrate that into an LLM, that is something that we really want to do.''} 

The challenge of data accessibility compounds these representation issues. I16 from Masakhane describes how African datasets were historically collected but remained inaccessible to African researchers due to cost barriers: \textit{``Prior to this community movement, many African datasets have been collected and Africans do not have access to them and have to pay... Their universities never prioritise [access] because it is too expensive,, so eventually they do not do much research. Eventually, we decided to address this by creating a dataset.''}

\subsubsection{Mentorship and skills development}

\noindent  Lowering barriers to AI research and creating mentorship opportunities, particularly for students, represents a key incentive for organisations like Masakhane. They promote data collection and language skills as valuable contributions while providing comprehensive mentorship, including programming, model training, GPU usage, experimental design, and academic writing. Several of their mentees have subsequently pursued graduate degrees abroad, demonstrating the program's effectiveness in developing AI talent.

\subsubsection{Public accountability for publicly funded research}

\noindent  For publicly funded R\&D projects or labs, releasing outputs to the public represents both an ethical obligation and a practical imperative. For example, I15 from the National Library of Norway explains their principle: \textit{``We have the principle that we release whatever we are allowed to release here. We are paid by the government, and so we should do that.''} Similarly, I17 from Fraunhofer IAIS emphasises that because the compute resources used to train Teuken-7B were funded by taxpayers (800,000 GPU hours for training plus additional experimentation hours), \textit{``as this is taxpayers' money, it is good to give something back to the public.''} 

\subsubsection{Peer recognition}
\noindent  Recognition motivates both individuals and organisations. I4 notes that open source communities and labs release models \textit{``to get recognition and to shape the narrative around who is doing the best things and who is on top of their game.''} Individual recognition is also important for career development for individuals in academia and industry alike.

\subsubsection{Personal passion and hobby}
\noindent  Personal passion for open source motivates many individuals. For example, I14 from Ant Group comments that, among the Chinese open source AI community, \textit{``I think 50\% are PhD students with full passion and with 16 hours a day they just sleep and work.''}

\subsection{Economic Motivations}

\subsubsection{Ecosystem building for competitive advantage}

\noindent  Open source provides a strategic approach for building ecosystems where diverse stakeholders pool resources and avoid duplication to compete effectively with leading AI companies. This collective approach is particularly important for smaller organisations and regions lacking the individual capacity to compete with major technology companies. For example, I1 from AI Singapore explains their ecosystem strategy of building foundations that are \textit{``70\% complete,''} enabling others to develop the remaining 30\% based on their specific needs. This approach allows smaller organisations to collaborate, pool resources such as compute and data, enable model reuse, and avoid duplicating effort.

Regional ecosystem building also represents a crucial strategy for areas lacking major AI capabilities. For example, I11 from SCB 10X explains: \textit{``we try to open source things because we want to foster innovation within Thailand and help the community grow, so that we can catch up with all the big tech players... We feel like we do not have the capacity to compete with, say, OpenAI.''} I17 from Fraunhofer IAIS emphasises this need in the European context: \textit{``in Europe we do not have companies that are as big in numbers as DeepSeek, let alone the US companies. So what we need in Europe instead is an ecosystem... Without open source, these ecosystems have no chance to grow fast or provide economic value.''}

The BigScience Workshop demonstrates ecosystem benefits through the mobility of its contributors, who subsequently brought code, infrastructure, and techniques developed during the workshop to other organisations, such as Mistral and Cohere. The workshop also created direct benefits for Hugging Face as the coordinating organisation, with I4 explaining that learnings from the collaboration shaped \textit{``how the hub evolved, how the datasets library evolved, how the inference libraries, open source inference libraries that we are putting out, evolved.''}

\subsubsection{Resource efficiency through model reuse}

\noindent  Economic considerations drive researchers and developers to derive models from existing LLMs rather than training new LLMs from scratch, enabling focused resource investment in fine-tuning for specific contexts not well covered by existing models. This approach is particularly valuable for organisations targeting specific languages or domains. For example, I11 from SCB 10X explains, \textit{``[The project] is looking for a solution that basically would be cheaper, so more cost-effective, and also more customizable. When it comes to open source models, you can customise them a little more.''} Organisations like AI Singapore, SpeakLeash Foundation, and Typhoon leverage this approach to achieve specialised capabilities in Southeast Asian languages, Polish, and Thai, respectively. 

Base model selection involves careful consideration of technical advantages for target applications. AI Singapore selected both Llama 3 and Gemma 2 models for complementary benefits, with I2 explaining that while Gemma better suits Southeast Asian languages due to the tokeniser and data advantages, Llama offers superior ecosystem interoperability. Similarly, Cohere for AI chose Google's MT5 model for Aya 101 because of its \textit{``broad language coverage''} across 100 languages, while SpeakLeash Foundation selected Mistral base models because they could achieve strong Polish language performance through fine-tuning.

\subsubsection{Gaining recognition for expertise}

\noindent  Gaining recognition for expertise is an important incentive for both companies and research institutes, enabling them to demonstrate capabilities and attract talent or funding. For example, I4 from the BigScience Workshop notes that open source participation allows developers and organisations to \textit{``get recognition and shape the narrative around who is doing the best things and who is on top of their game.''} Similarly, I7 from BAAI describes how their open source strategy helps recruit top talent who might otherwise prefer to work in industry. Furthermore, for regions lacking major technology companies, this recognition becomes particularly valuable. I12 from SCB 10X explains: \textit{``a lot of parties in Thailand from industry who would like to be perceived as a leader in the tech space here... if you only have a proprietary, internal model, no one will know how good your model is. People will not talk about your stuff.''} Meanwhile, I14 from Ant Group describes corporate motivations to \textit{``make their company famous because they contribute to the open source project [hoping] the project will mention their company name on the next release.''} 

\subsubsection{Career advancement}

\noindent  Career advancement motivates researchers and developers, as working openly enables transparency and promotion of one's abilities to developer communities and potential employers. For example, I16 from Masakhane explains that, for academic researchers, open source and open science facilitate the use and citations of their work, which is key to academic career progression.

\subsubsection{Business strategies (e.g., market entry, ecosystem expansion)}
\noindent  Companies leverage open LLMs as strategic tools for market entry and ecosystem expansion. For example, I6 from Meta explains that they promote Llama models across third-party devices and platforms ranging from augmented reality glasses and mobile phones to medical devices. 

\subsection{Technological Motivations}

\subsubsection{Promoting open science and reproducibility}

\noindent  Interests in open science and reproducibility drive many open LLM projects, as AI researchers seek to recreate powerful closed-source models in transparent, reproducible ways. EleutherAI, the BigScience Workshop, and OpenGPT-X initially aimed to recreate OpenAI's GPT-2 and GPT-3 models. For example, I17 from Fraunhofer IAIS explains that he and his supervisor initiated their grant proposal for the OpenGPT-X project in 2021 to build an open, reproducible GPT-3 based on their scientific interest. Similarly, Hugging Face's SmolLM team and I14's lab at Ant Group sought to recreate DeepSeek's R1 model capabilities. I14 from Ant Group explains that DeepSeek's R1 release served as a major motivator to reproduce the results and openly share the ingredients for doing so, particularly since DeepSeek \textit{``only open sources its model weights, but not the datasets and the training frameworks.''} 

\subsubsection{Standardization}

\noindent  Establishing open standards for LLM development enables researchers to build upon each other's work effectively. For example, I3 from EleutherAI and I9 from Hugging Face explain that simplifying LLM pre-training motivated their development of the GPT-NeoX and nanotron libraries, respectively. I9 emphasises that, \textit{``It would be nice if there was one pre-training framework everyone contributed to, but internal use cases and motivations often conflict. Some prefer production-ready frameworks, even if they come with more bloat, while others want lightweight frameworks with minimal features for easier experimentation. Hopefully, we can converge someday on a balanced solution.''} The same motivation goes for standardising evaluation tools. I7 from BAAI emphasises their inability to evaluate models for every specific use case or domain, underscoring the need for community-developed benchmarks and evaluation frameworks. I3 from EleutherAI and I9 from Hugging Face also cite the ability to run systematic, reproducible LLM evaluations as a key motivation for developing their LM Evaluation Harness and lighteval frameworks, respectively.

\subsubsection{Demonstrating the capabilities of small models}

\noindent  Advancing research on small models and proving their competitive capabilities against larger models is also a motivation. For example, I9 from Hugging Face explains that SmolLM aimed to \textit{``advance research on small models and prove that small models can compete with larger models, while being cheaper to operate.''} 

\section{Governance and Community Engagement Approaches in Open LLM Projects (RQ3)}\label{sec:results-governance}

\begin{framed}
\noindent $\diamond$ \textbf{Summary:}
\noindent We identify five organisational models in open LLM projects: \newline 

\noindent $\diamond$ In \textbf{company-led projects} (e.g. Meta's Llama, Hugging Face's SmolLM), a single company typically maintains centralised control of the LLM development process and may engage in selective collaborations with external stakeholders (e.g., in order to access specific expertise) prior to the release of the open LLM. \newline 

\noindent $\diamond$ Research institute projects include \textbf{single research institute projects} (e.g., SEA-LION, OLMO, Aquila) and \textbf{multi-organisational research institute projects} (e.g., OpenGPT-X). \newline 

\noindent $\diamond$ Grassroots projects encompass \textbf{non-profit-sponsored grassroots projects} (e.g., Pythia, Bielik, Aya) as well as \textbf{company-sponsored grassroots projects} (e.g., the BigScience Workshop). They utilise hybrid governance combining centralised project coordination with decentralised development and contributions.
\newline
\newline
Community engagement strategies vary across these five organisational models and the stages of the open LLM lifecycle, with less collaboration during the pre-release stages due to structural constraints in LLM development. Platforms like Discord, X, Slack, and Hugging Face Hub are commonly used for knowledge sharing and collaboration across distributed teams and developer communities.
\end{framed}

\subsection{Organisational Models and Governance Approaches in Open LLM Projects}

\begin{figure*}[h]
    \centering
    \includegraphics[width=0.7\linewidth]{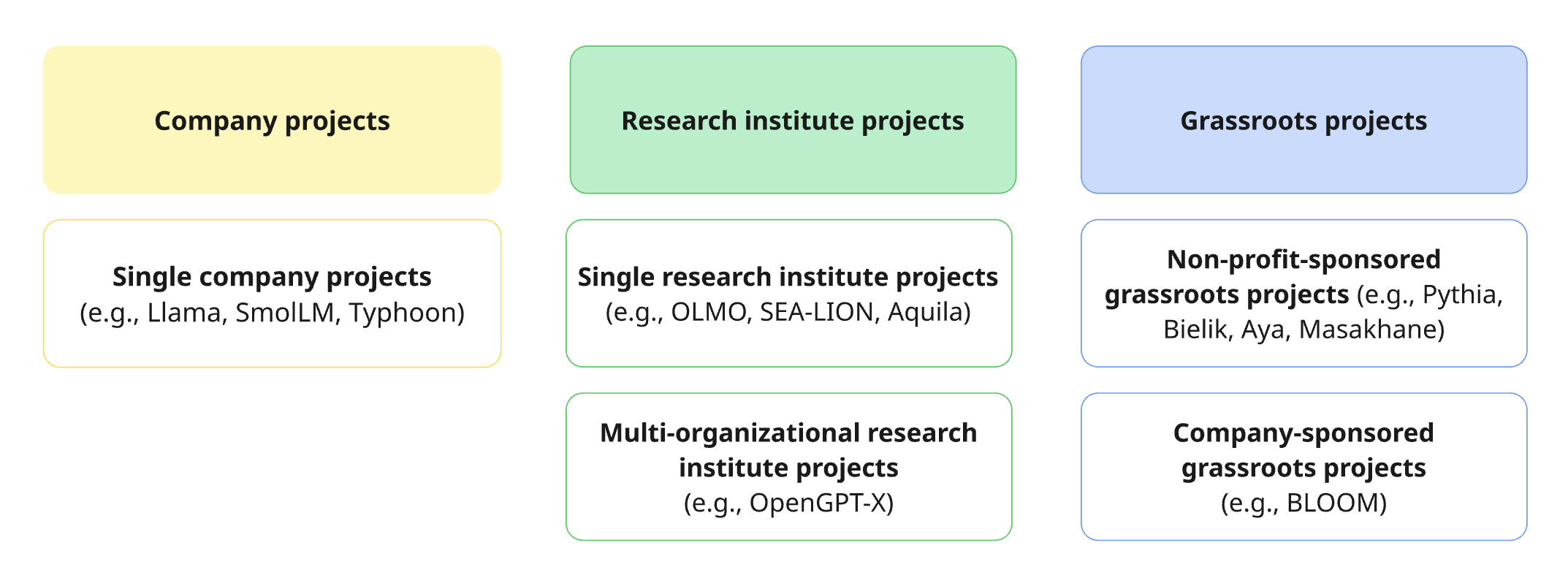}
    \caption{Organisational models of open LLM projects}
    \label{fig:governance}
\end{figure*}

\noindent  Open LLM developers coordinate their development processes using governance frameworks that reflect their institutional origins, resource constraints, and community engagement philosophies, including centralised control, distributed community-driven development, and hybrid approaches. We identify five organisational models: single-company-led projects, single-research-institute projects, multi-organisational research-institute projects, non-profit-sponsored grassroots projects, and company-sponsored grassroots projects, as illustrated in Fig.~\ref{fig:governance}.

\subsubsection{Company-led projects}
\noindent  In single-company-led projects, companies typically develop open LLMs through internal processes, with limited external collaboration during development. LLMs are used and embedded in commercial products and services, thereby becoming core components of their value offerings. The LLMs can also serve as a strategic tool for growing adoption through user communities and business ecosystems, thereby driving both innovation and business value upstream to companies. In these cases, the open LLM projects can be considered single-company-led projects, similar to single-company or vendor-sponsored OSS projects. 

Meta and Hugging Face exemplify this centralised approach with their Llama and SmolLM projects, respectively, which involved minimal external collaboration before the models' release. Companies (or vendors) distribute decision-making power and authority, typically among their internal or contracted staff. With this authority, they make decisions and maintain a final say on the design and implementation of the LLMs. The communities form a user community around the base models, extending and enriching them through post-release reuse and targeted modifications in derivative models (e.g., for specific languages or domains). The communities can make requests and provide input through the communication channels and development platforms leveraged by the projects, which the single company can, in turn, choose to consider and respond to. 

Control of downstream use and any derivative models can be maintained through the company's licensing schema. For example, in the case of Meta's Llama models, released under the restrictive Llama Community License Agreement, the Llama trademark must be maintained in derivative model naming conventions, and users above certain thresholds require special licenses. In essence, this limits the user community's freedom while still enabling their adoption and development of the base models under the current license terms.

In some cases, companies initiate strategic collaborations with external stakeholders to gain access to specific expertise. For example, I11 from SCB 10X describes their partnership approach as existing \textit{``on a spectrum''} from formal agreements to casual collaborations where they \textit{``jump in and help them.''} While their internal team handles most core work, they selectively engage external collaborators for specific tasks: \textit{``We do outsource a couple of people to help us on data collection and annotation as well and usually like other researchers who come help...There will be, like, an academic researcher at a Thai university. We do have connections with them and we can ask them to come help.''} Such selective engagements allow companies to maintain control while accessing specialised expertise when needed.

\subsubsection{Research institute-led projects}

\noindent  Among research institutes, two governance models stand out among the sampled projects: single research institute and multi-organisational research institute open LLM projects. 

\textbf{Single research institute projects.} In single research institute projects, a main research institute drives and has authority over the LLM development process, exemplified by Ai2 and AI Singapore. Still, there is a high reliance on partners and donors to enable the research and model development, which extends soft power and influence beyond the single institutes. Collaboration with external partners can be characterised by bureaucratic constraints and formal procedures. For example, I1 from AI Singapore explains that, as they are a project at the National University of Singapore, onboarding external contributions involves \textit{``some bureaucracy involved in trying to formulate the scope of the partnerships''} with most contributions requiring contractual agreements. They must be \textit{``clear about what they're able to contribute or not...so that there is documentation for all of this so that everything is above board, especially when it comes to the fact that we are hosted by an academic institute that is renowned internationally.''} However, partnerships often exceed formal agreements: \textit{``A lot of the people who work with us end up contributing so much more than what is in the collaboration agreement, just because they want to do something good for the region.''}

\textbf{Multi-organisational research institute projects.} Multi-organisational research institute projects bring several stakeholders together, typically through joint research and grant-funded projects. I15 from the National Library of Norway explains that collaborations are \textit{``definitely driven through this kind of common financing''} from sources like the Norwegian Research Council or EU partnerships. Their neutral institutional position facilitates collaboration between competing organisations: \textit{``as a neutral actor, competing organisations have collaborated in quite a few projects through us since we are sort of in the middle and we cannot sort of favour''} one party over another. The OpenGPT-X project is also a formal multi-partner collaboration, involving 10 German organisations, including research centres, universities, and companies, with clearly defined work packages covering infrastructure optimisation, data filtering, training, and evaluation. It is led by Fraunhofer IAIS, and each partner had specific roles determined by the grant agreement, such as TU Dresden handling model evaluation and checkpointing procedures. The distribution of authority and decision power, and execution of it, is therefore typically managed through joint partnership agreements. 

\subsubsection{Grassroots organisation-led projects}
Grassroots organisation-led projects have used hybrid governance approaches that combine centralised coordination of critical infrastructure and decision-making with decentralised community-driven development. Among the sampled projects, two models stand out: non-profit-sponsored and company-sponsored projects.

\paragraph{Non-profit-sponsored grassroots projects.}
\noindent  In non-profit-sponsored grassroots projects, decision-making and development authority are maintained by a core group of people who are hired or closely affiliated with a non-profit organisation. The goal is typically to ensure the project's overall vision, quality assurance, and general development productivity, while still enabling community engagement and active collaboration. 

EleutherAI operates with dual governance modes depending on project criticality, with I3 explaining, \textit{``I think that there isn't one governance to rule them all, but that governance structures and strategies are tools to accomplish your needs.''} For widely used tools like GPT-NeoX and LM Evaluation Harness, the core team maintains centralised control to ensure \textit{``design consistency''} and avoid \textit{``breaking previous implementations''} for the thousands of production users. I3 explains, \textit{``We value community input... but that at the end of the day, that we are the decision makers... because our two big libraries that a lot of people are using are used in production by dozens in the case of the training library and thousands in the case of the evaluation library.''} Simultaneously, EleutherAI supports bottom-up community projects through dedicated Discord channels where contributors self-organise independent initiatives. They also enable community leadership of official projects when staff bandwidth is limited: \textit{``People come to us and say, `Hey, this would be super cool. I think you guys should do it.' And we are like, `We'd love to do that. We do not have someone who can be in charge of this right now. We are bandwidth-constrained. Would you like to be in charge of it?'''} I3 adds that their non-profit status provides flexibility that companies lack, as they are not concerned with \textit{``building their ecosystem control or increasing others' dependency on them.''}

The SpeakLeash Foundation operates in a similar way to empower its community, including over 3,500 passive and 100 active Discord members. While project founders make critical decisions about training their Bielik models, the community operates in a decentralised manner for other initiatives, such as educational adaptations. Their community created tools for data collection, quality assessment workflows, and applications for gathering human preferences, while contributing over 60,000 responses to toxicity assessment surveys.

Another example of a non-profit-sponsored grassroots project is that of Aya. Cohere for AI, the non-profit arm of the startup Cohere, positions itself as \textit{``initiators''} while empowering community contributors to shape project direction. Their Aya project involved 3,000+ contributors, with I5 explaining that contributors from under-resourced languages showed particular passion: \textit{``For many low resource languages, it is easier to collect data for because the people who speak those languages, they know it is low resource... they've experienced it not being included in any language model. So, they are really passionate to contribute.''} Contributors could determine their involvement level and leadership roles, with the Cohere for AI team providing structure, pacing, and deadlines while maintaining that projects are \textit{``clearly built by the community for the community.''}

\paragraph{Company-sponsored grassroots projects.} Company-sponsored grassroots projects differ from single company-led projects in that governance is more open and decentralised, while the sponsoring company typically has a big influence and, on some occasions, the final say. A prime example is that of the BigScience Workshop. The project organised a community of over 1,000 volunteers from 250+ organisations into specialised working groups, though decision-making power remained concentrated with Hugging Face as the coordinating organisation. While working groups covered areas like data, modelling, and organisation, most Hugging Face employees participated, with 12-45 people working full-time on the project. Only Hugging Face interacted with compute administrators, and their Chief Scientific Officer held final decision authority. However, this committee approach faced efficiency challenges. I3 from EleutherAI comments: \textit{``BigScience was done by committee, and it left me with the impression that that was not a great way to govern a research project. It took a lot of work, a lot of time. Committees are very inefficient ways to reach consensus...when you're trying to kind of reach a decision that can guide the behaviour of dozens of people, it is not really about finding agreement. It is about finding compromise.''} This experience suggests that, while inclusive, committee-based governance can create coordination overhead that may limit project efficiency. 

\subsection{Community Engagement Strategies and Platforms}

\subsubsection{Platform-mediated community building}

\noindent  Different platforms serve distinct functions for communication and collaboration in the open source AI community, from real-time discussion and knowledge sharing to asynchronous collaboration and resource distribution. The choice of platforms impacts community dynamics and collaboration effectiveness.

Discord is a popular community hub and knowledge-sharing platform that enables real-time interaction and the sharing of expertise across diverse contributor communities. EleutherAI's Discord server functions as a \textit{``hub for conversation about AI technology,''} where, as I3 explains, experts \textit{``join in on conversations regularly and help to answer questions,''} including \textit{``developers from otherwise closed-source companies, as well as people who are just interested in AI technology.''} This creates valuable knowledge-sharing opportunities: \textit{``They're able to come talk about it with other researchers and this like mostly public environment, which is, I think, really great because, you know, we do not want knowledge to stay locked up.''} Similarly, I17 from Fraunhofer IAIS notes that Discord enables more \textit{``lively scientific discussions''} compared to other platforms, facilitating community sharing of model modifications and optimisations.

Slack is also widely used for coordination and partnerships, providing tools for communication and coordination in working groups and inter-organisational collaborations. For example, the BigScience Workshop used Hugging Face Slack for `\textit{`conversations and meeting coordination within and across working groups,''} with I4 explaining that it enabled \textit{``working with people in different time zones''} and remains a site for ongoing collaboration initiation.

Multi-platform coordination is also common in projects spanning multiple regions and cultures. For example, Cohere for AI managed global coordination challenges through platform diversity, with I5 explaining they \textit{``often held the same meeting in four different time zones''} and relied on Discord for asynchronous coordination and resource sharing. However, recruiting community leaders required adapting to local preferences: community leads for their 101 target languages \textit{``preferred a range of online platforms, including WhatsApp.''}

\subsubsection{Community recruitment and growth strategies}

\noindent  Targeted and open recruitment approaches enable projects to balance expertise needs with inclusive participation. The BigScience Workshop employed both strategies, starting with 200-300 targeted invites followed by open invitations that attracted around 1,000 participants. I4 explains how they sought to expand their recruitment efforts beyond their personal networks: \textit{``I definitely spent a lot of time trying to look beyond that network for people who might be relevant, and then sending to the extent possible customised [invitations]... also trying to make sure to push in some of the directions that we were missing. So, for example, 'okay, so who is going to be interested in South America, who is going to be interested in if you're going to add African languages, who is going to be interested in working on that?'''} They also recognised that \textit{``expertise comes in several forms''} and emphasised that researchers could make both technical and non-technical contributions to the project.

Fostering a welcoming culture is crucial for organic community growth and sustaining community-driven projects. Cohere for AI's community spans from \textit{``assistant professors in London to high school students in India,''} maintained through welcoming community norms. I5 explains their approach: \textit{``I think one of the main steps is just the creation of this Discord community and also how we are maintaining that community, so it is a very friendly, open community. There is no room really to be judgmental if you do not know something...we really have tried to set this tone and our community members are great at that.''} This culture enables bottom-up initiatives where \textit{``connections form there can lead to research down the road.''} 

As project communities scale, it becomes valuable to hire a full-time community manager to animate the community. I16 from Masakhane explains their decision to \textit{``hire a community manager to be in charge of community and mentorship management and communications because previously they were reliant on volunteers, but volunteers move on due to life or job changes, and they want to continue running Masakhane as a community initiative.''} 

\subsubsection{Community outreach and user engagement}

\noindent  Showcase-driven engagement strategies are one way open LLM developers demonstrate their LLMs' capabilities and increase adoption of their models and tools. For example, the Typhoon developers organized a \textit{``Typhoon Application Week''} to \textit{``rally/grow the community and showcase applications.''} Rather than directly promoting usage, they \textit{``have a few examples of how we can use Typhoon in our workflow...just to kind of, like, inspire people or show people, hey, we can start using Typhoon in your workflow, in your applications.''} This approach builds community engagement through demonstration rather than direct marketing.

Structured feedback collection and user support also enable open LLM developers to maintain relationships with their user communities while gathering insights for improvement. AI Singapore created a Developer Advocacy team to \textit{``identify user needs and collect feedback in a more structured way and expand the scope of the project beyond academic and technical aims.''} They provide early model access to partners for testing and feedback while accommodating community requests for different formats and deployment environments. I6 from Meta estimates their community size through platform metrics, noting \textit{``650 million downloads now of Llama models, derivatives, and there is like 60,000...basically, what I would call forks, which are like models that have been finetuned and then re-uploaded.''} 

Collaborative events and networking create opportunities for relationship-building within and beyond the project's community. I15 from the National Library of Norway describes participating in Hugging Face hackathons: \textit{``Those have been very useful for us...You learn a lot through those hackathons, right? But also because we get to know the people, both at Hugging Face, but also people from Sweden and Denmark, for instance, with whom we collaborated later. So those have been great meeting points.''} These events serve dual purposes of skill development and network building that enable future collaboration opportunities.

\section{Discussion}\label{sec:discussion}

\subsection{Implications for Research}
\noindent To synthesise our findings, we propose a conceptual model of open collaboration in open LLM ecosystems, summarised in Table~\ref{tab:conceptualModel}. Our findings reveal that open collaboration on open LLMs extends far beyond the models themselves, happening across distinct artefact domains and is shaped by domain-specific properties that influence how participation is structured and distributed. At the same time, collaboration is not static, but evolves across lifecycle stages as different domains become more prominent. These dynamics are further influenced by diverse developer motivations—--from democratising AI access and development to building competitive ecosystems—--and are coordinated through a range of governance structures, from single-company-led initiatives to non-profit-sponsored grassroots projects. Taken together, the model captures how collaboration in open source AI emerges from the interaction between artefact domains, their underlying properties, lifecycle dynamics, motivations, and governance arrangements.

\begin{table*}[]
\caption{Conceptual model of open collaboration in open LLM ecosystems, summarising the key constructs, including artefact domains, domain properties, lifecycle stages, motivations, and governance structures.}
\label{tab:conceptualModel}
\centering
\begin{tabular}{p{3cm}p{14cm}}
\toprule
\multicolumn{2}{l}{\textbf{Artefact domains}} \\ \midrule
Model & Development, training, and sharing of model architectures and weights, typically requiring specialised expertise and coordinated effort. \\
Data & Collection, annotation, and curation of training and finetuning datasets, often labour-intensive and constrained by ownership and quality considerations. \\
Software & Development of open source training, inference, and evaluation frameworks that enable and shape how models are built and deployed. \\
Evaluation & Creation of benchmarks, leaderboards, and testing practices that guide model improvement and enable comparative assessment. \\
Non-technical & Documentation, community management, and knowledge sharing that support coordination, onboarding, and sustained ecosystem engagement. \\
Compute & Provision and access to computational infrastructure through partnerships or shared resources, enabling large-scale training and experimentation. \\ \midrule
\multicolumn{2}{l}{\textbf{Domain properties}} \\ \midrule
Participation Access & The degree to which actors are able to take part in a domain, ranging from restricted access in core domains to broad accessibility in community-facing ones. \\
Expertise Requirement & The level and type of knowledge needed to contribute effectively, from highly specialised technical expertise to more general and transferable skills. \\
Modularity & The extent to which contributions can be made independently of the broader system, enabling either tightly coupled or distributed forms of work. \\
Control \& Ownership & The extent to which artefacts are governed by legal, organisational, or institutional constraints limiting openness and reuse. \\ \midrule
\multicolumn{2}{l}{\textbf{Life-cycle Stage}} \\ \midrule
Pre-training & Collaboration is concentrated in core domains, relying on selective partnerships, specialised expertise, and access to compute infrastructure. \\
Post-training & Collaboration expands selectively beyond the core, with controlled sharing of models and datasets enabling targeted contributions. \\
Post-release Reuse & Collaboration becomes broadly distributed through platform-mediated interactions, enabling fine-tuning, evaluation, and derivative development. \\ \midrule
\multicolumn{2}{l}{\textbf{Motivational Drivers}} \\ \midrule
Social & Motivations related to democratising AI, increasing language representation, and addressing broader societal and equity concerns. \\
Economic & Motivations tied to ecosystem building, cost sharing, and leveraging open collaboration for competitive advantage and network effects. \\
Technical & Motivations focused on advancing open science, improving model capabilities, and standardising tools and practices. \\ \midrule
\multicolumn{2}{l}{\textbf{Governance Structures}} \\ \midrule
Single company-led projects & Centralised control by a single firm, with selective external collaboration to access complementary resources and expertise. \\
Research institute projects & Institution-led initiatives, balancing openness with funding and partnership constraints shaped by public or academic governance. \\
Multi-institute projects & Structured collaboration across multiple organisations with defined roles, coordination mechanisms, and shared objectives. \\
Non-profit-sponsored grassroots projects & Hybrid models combining central infrastructure control with decentralised, community-driven contributions. \\
Company-sponsored grassroots projects & Company-coordinated initiatives that mobilise large volunteer communities while maintaining strategic oversight. \\
\bottomrule
\end{tabular}
\end{table*}

\paragraph{Collaboration takes place across multiple artefact domains, not just on models.} We find that collaboration in open LLM projects extends beyond the models themselves, encompassing collaborations on interdependent artefacts that are crucial for developing and evaluating LLMs, including training and finetuning datasets, open source training and evaluation frameworks, open benchmarks and leaderboards. More formally, we identify six artefact domains for collaboration as visualised in table~\ref{fig:framework}: models, data, software, evaluation, non-technical, and compute. 

These domains capture distinct, but interrelated, loci of collaborative activity. Model collaboration centres on the development, training, and sharing of model architectures and checkpoints. Data collaboration involves the collection, annotation, curation, and dissemination of datasets for both pre-training and post-training. Software collaboration focuses on the development and maintenance of open source training, inference, and evaluation frameworks that shape how models are built and deployed. Evaluation collaboration encompasses the creation of benchmarks, leaderboards, and testing practices that enable comparative assessment and guide model improvement. Compute collaboration relates to access to and provisioning of computational resources, often through partnerships or shared infrastructure, which are essential for large-scale training and experimentation. Finally, non-technical collaboration includes documentation, community management, knowledge sharing, and user feedback, which support coordination, onboarding, and sustained engagement across the ecosystem.

This finding challenges narrow, model-focused views of open source AI, suggesting that research on open collaboration in open source AI requires a broader perspective that considers how different artefacts and stakeholders interact throughout the development and reuse lifecycle of open LLMs. 
This finding further extends Choksi et al. ' s~\cite{choksi_brief_2025} argument that the open source AI landscape should not be viewed as a monolith but rather as a fragmented constellation of both ephemeral and sustained developer communities with distinct expertise and interests, ranging from technical optimisation to domain- or language-specific applications.

\paragraph{Domain constraints shape opportunities for collaboration}
Our findings further suggest how these domains can be characterised with respect to how collaboration is organised and who is able to participate. 

First, domains vary in the accessibility of participation; i.e., how easy it is for actors to contribute and participate in the collaboration. Domains such as model development and computing are typically restricted due to dependencies on infrastructure, coordination, or organisational control, whereas evaluation and non-technical domains tend to be more collaborative. 

Second, the domains differ in the type and level of expertise required, ranging from highly specialised, often tacit knowledge in model development and alignment to more widely accessible or learnable skills in areas such as documentation, application development, and benchmarking. Some domains may be open yet require substantial expertise, while others are restricted despite relatively modest skill requirements.

Third, domains vary in modularity, meaning the extent to which contributions can be made independently of the broader open source AI system. Highly modular domains, such as training and evaluation frameworks, enable contributors to work on relatively self-contained components, facilitating parallel and distributed development. In contrast, model and data domains are more tightly coupled, requiring integration with complex pipelines and coordination with core development processes. Fourth, domains also differ in control and ownership, reflecting the degree to which contributions are governed by legal, organisational, or institutional constraints. Data and compute are often subject to strong ownership and access restrictions, whereas evaluation and non-technical domains are more frequently governed by open, community-based practices. 

Finally, domains play different functional roles in the ecosystem, distinguishing between core domains (models, data, and compute), which are directly required for system development, and enabling domains (software, evaluation, and non-technical), which support, coordinate, and extend the AI system use and evolution.

Together, these aspects influence and help explain the opportunities and challenges for collaboration across the different collaboration domains, e.g., why evaluation, software, and community engagement scale more broadly, while others remain concentrated among a smaller set of actors. In this sense, openness in open source AI is not a monolithic characteristic but an emergent property of how collaboration is organised across different yet interconnected domains.

\paragraph{Collaboration patterns vary by lifecycle stage} 
We identify distinct collaboration patterns across LLM development stages that differ significantly from traditional OSS development practices. Rather than remaining relatively stable over time, collaboration in open LLM projects evolves as different artefact domains become more or less prominent across the development lifecycle.

During pre-training, collaboration is predominantly concentrated in core domains—models, data, and compute—where low participation access, high expertise requirements, tight coupling, and strong control over resources lead to selective forms of collaboration. This is reflected in strategic partnerships (e.g., AI Singapore with Meta and Google), specialised expert contributions (e.g., contributions to the mixture-of-experts architecture for EleutherAI’s GPT-NeoX framework), and infrastructure sharing (e.g., the BigScience Workshop’s access to the French government’s Jean Zay supercomputer). In this stage, collaboration is limited to actors capable of engaging with tightly coupled and resource-intensive artefacts.

The post-training stage exhibits a partial shift, where collaboration expands into additional domains but remains selective. While core domains still play a central role, increased modularity and reduced integration complexity enable more targeted forms of participation. This is reflected in the controlled sharing of model weights with trusted partners (e.g., Hugging Face’s sharing of SmolLM weights with startups) and specialised dataset releases (e.g., BAAI’s Infinity Instruct dataset). Collaboration in this stage reflects a transitional regime, where contributions are possible beyond the core, but remain gated by expertise, coordination, and control considerations.

In contrast, the post-release stage is characterised by a marked expansion of collaboration into enabling domains, where higher participation, access and greater modularity support broad community engagement. Platform-mediated interactions enable a wide range of contributions, including model fine-tuning (e.g., GoTo’s Javanese assistant based on SEA-LION), evaluation contributions (e.g., AfroBench’s crowd-sourced benchmark expansion), and derivative development (e.g., the community-built Maya derived from Aya models). In this stage, contributors can engage through more loosely coupled artefacts without requiring direct involvement in core model development.

This lifecycle variation represents a fundamental difference from typical OSS projects, where collaboration structures tend to remain relatively consistent across development stages. To speak in Raymond’s terms \cite{raymond_cathedral_2001}, whether an open LLM initiative is developed with a “cathedral” or “bazaar” approach varies less between initiatives than within them across the lifecycle, with pre-training and post-training stages generally exhibiting cathedral-like characteristics due to the properties of core domains, and post-release stages increasingly reflecting bazaar-like, distributed collaboration enabled by more accessible and modular domains.

\paragraph{Social, economic, and technological motivations driving collaboration in open source AI} Open LLM developers have various motivations, including social motivations (e.g., democratizing AI, expanding language representation), economic motivations (e.g., ecosystem building, resource efficiency), and technological motivations (e.g., promoting open science, standardising tools). The prominence of social motivations — particularly around democratisation and representation — reflects the societal stakes of AI development, where concerns about the concentration of power and technological sovereignty have become central to AI policy debates. Meanwhile, the economic motivations we observe suggest that open collaboration in AI is strongly intertwined with commercial strategies, as organisations recognise that open ecosystems can accelerate innovation, reduce costs, and establish competitive moats through network effects and ecosystem lock-in.

\paragraph{Diverse yet professionalised governance frameworks and organisational models dominate in open LLM projects} We identify 5 distinct governance models in open LLM projects: single company-led projects, research institute projects, multi-institute projects, non-profit-sponsored grassroots projects, and company-sponsored grassroots projects.

In company-led projects, a single company maintains centralised control throughout the pre-release LLM development stages and may engage in strategic external partnerships to gain access to external expertise or resources. 
While the single-company model has received attention in the OSS context~\cite{riehle2012single}, open source AI introduces further nuances that warrant consideration, e.g., how to grow and engage user communities across the various artefacts pre- and post-release. Licensing practices and business model design are important areas for future work, exploring how they can support an open innovation and collaboration process while ensuring business value for the company.

Public research institutes have used single-institute and multi-organisational governance models. 
While similarities can be drawn to both single- and multi-company governance models in OSS~\cite{schaarschmidt2015firms}, there are fundamental differences, e.g., in what incentivises the organisations (e.g., science vs. profit), and how novel and valuable information is managed (e.g., open science vs. corporate trade secrets). Open collaboration may, hence, have larger potential in the context of projects operated by public research institutes. Yet, as these are governed by their respective funding sources and complex partner and project agreements, openness can be stifled quite easily. Future work is needed to shed light on how open collaboration can be incentivised, enabled, and carried out across the public research context.

Grassroots projects employ hybrid governance models that combine centralised coordination with decentralised community development. Non-profit-sponsored projects like EleutherAI maintain dual governance modes—centralised control over critical tools and infrastructure (e.g., GPT-NeoX, LM Evaluation Harness) while supporting bottom-up community self-organisation (e.g., through their Discord channel). Company-sponsored grassroots projects like the BigScience Workshop and Aya were coordinated by lead sponsor companies—Hugging Face and Cohere,, respectively —but involved large volunteer communities.
These grassroots models are the ones that show the largest potential for open collaboration across our surveyed cases, and provide inspiration for both current and future open source AI projects, similar to community-driven OSS projects~\cite{alrawashdeh_user_2020}. Future research may support the open source AI ecosystem by further expanding on how governance and coordination practices can be designed and function at both small and large scales. Complexity further increases as consideration is required for how the peer-production model can accommodate intricate dependencies across data, software, and model collaborations and projects.

While our analysis focused on open LLM project governance models, we note that, apart from copyright concerns and data licensing, the topics of responsible AI practices and safety considerations are largely absent in the findings. We note that this absence is striking, given the growing policy focus on AI safety governance, especially as open models approach parity with proprietary systems in reasoning, agentic capabilities, and multimodal generation. We recommend future research to investigate how open LLM developers are navigating AI safety considerations.

\paragraph{Community engagement platforms} Platforms like Discord, X, Reddit, Slack, and Hugging Face Hub are popular platforms for knowledge sharing, discussions, and collaboration across geographic and organisational boundaries, with each serving distinct functions. Discord is widely used to facilitate knowledge sharing and community building (e.g., EleutherAI's server became a \textit{``hub for conversation about AI technology''} where experts help answer technical questions). Slack supports formal project coordination for structured collaborations. For example, in the BigScience Workshop, Slack enabled \textit{``conversations and meeting coordination within and across working groups''} across time zones. Hugging Face Hub serves as the primary artefact sharing platform, functioning as \textit{``the de facto way to share your models and to make things accessible to people.''} Platform choices also reflect cultural preferences. Cohere for AI discovered that community leaders across their 101 target languages \textit{``preferred a range of online platforms, including WhatsApp,''} requiring multi-platform coordination strategies. Given the distributed, multi-platform nature of open LLM collaboration, future research should adopt multi-platform approaches to capture the full scope of community interactions and coordination practices that span these diverse platforms.







\subsection{Recommendations for Practitioners}

\noindent  Based on the interviews, we provide the following recommendations, presented in no particular priority order, for stakeholders seeking to support the global community of practitioners building a more open future for AI.

\subsubsection{AI researchers and developers}
\begin{itemize}
    \item \textbf{Proactively engage with open source AI projects and communities.} Do not wait for collaboration opportunities to find you; actively reach out to potential partners and communities. As I11 from SCB-10X advises: \textit{``do not be shy and just reach out.''} Join Discord servers, participate in hackathons, and seek out opportunities to contribute to existing projects.
    
    \item \textbf{Avoid reinventing the wheel.} Before starting a new project, check if a solution already exists. As I12 from Typhoon emphasises: \textit{``Do not reinvent the wheel, a lot of knowledge is out there.''} Building on existing foundations enables faster progress, is more resource efficient, and can lead to new collaborations.

    \item \textbf{Experiment with  ``open lab'' approaches following the Marin model.} Experiment with ``open lab'' approaches that make the entire research process transparent by using OSS collaboration tools~\cite{hall2025marin}. 

    \item \textbf{Build evaluation datasets to support targeted improvements in underrepresented languages.} Invest in developing training and evaluation datasets for medium and low-resource languages that lack adequate representation in existing LLMs. 


    \item \textbf{Develop open source tools and best practices for AI safety.} As open models achieve near-parity with proprietary models, open source AI researchers and developers should develop and share tools and best practices for safe and responsible development, such as preventing malicious fine-tuning, verifying model integrity, and detecting misuse.
    
    \item \textbf{Develop full-stack AI expertise beyond fine-tuning.} I17 from Fraunhofer IAIS recommends mastering the entire AI pipeline \textit{``from data collection to post training...own your data collection, your data filtering, your data processing, how you filter the model, how you evaluate your model.''} This comprehensive understanding can enable unique contributions and reduce dependence on commercial providers.

    \item \textbf{Foster synergies between AI infrastructure providers, AI researchers, and industry.} I17 from Fraunhofer IAIS recommends fostering synergies between researchers, compute providers, and industry partners to accelerate the translation of research into innovation, learning from the example of the Swiss AI initiative.

    \item \textbf{Ensure fair compensation for inclusive participation.} I16 from Masakhane emphasises avoiding \textit{``extractive relationships''} by understanding entry barriers for potential contributors, providing mentorship opportunities, and compensating contributors fairly. For example, I16 recommends asking contributors how they want to be recognised or compensated to maintain their motivation and overall community health.
\end{itemize}

\subsubsection{AI companies}
\begin{itemize}
    \item \textbf{Invest in ecosystems rather than competing alone.} Particularly for companies outside major AI hubs, collaborative ecosystem development can collectively compete with industry giants. Pool resources for shared infrastructure, datasets, and standards while specialising in unique applications or regional needs.
    
    \item \textbf{Explore hybrid governance frameworks.} Learn from the grassroots project governance models, represented by EleutherAI, among others, which combine centralised control of critical production infrastructure by a core team while enabling community engagement and the infusion of new ideas in a bottom-up way.
    
    \item \textbf{Invest and collaborate on data provenance and licensing clarity.} While legal frameworks around training data usage may vary by jurisdiction (e.g., fair use or not), investing in clear data provenance can address practical and legal challenges faced by open LLM developers. 

    \item \textbf{Partner up and contribute to open source AI through compute donations.} Grassroots projects and research institute projects are often dependent on the donation and provisioning of computing to train models. By contributing compute resources, companies can both support and benefit from open LLM projects.
\end{itemize}

\subsubsection{Policymakers}
\begin{itemize}
    \item \textbf{Fund public AI infrastructure, not just models.} Public funding across the entire AI stack, including compute infrastructure, data curation, and OSS development and maintenance, can support the long-term sustainability of the open source AI ecosystem, while advancing crucial public interest areas, such as research, open source frameworks, and open benchmarks for AI safety.

    \item \textbf{Support public interest projects via compute donations or subsidies.} The examples of the French government's compute donation to the BigScience project and the crucial role of the Swiss National Supercomputing Centre in the Swiss AI initiative's development of the Apertus model highlight the role governments can play in funding the use of supercomputer infrastructure for the development of open LLMs that are designed for the public benefit. 
         
    \item \textbf{Support regional language and cultural representation through targeted initiatives.} Fund collaborative projects like AI Singapore's SEA-LION or Cohere for AI's Aya that engage native speaker communities in data collection and evaluation. This is especially relevant for public bodies representing cities, regions, or countries with diverse language communities, as it can ensure that the adoption of AI is paired with improvements in model capabilities that will benefit all these communities.
    
    \item \textbf{Ensure ``public money, public AI'' principles with appropriate licensing.} Require publicly funded AI research to use permissive licenses and release key artefacts (i.e., data, code, models, documentation) to maximise public benefit. 

    \item \textbf{ Bridge the developer-policy gap on AI safety governance.} Engage with open source AI developer communities to understand current safety best practices and support the development and adoption of open source safety tools, open benchmarks, and related resources, following the example of the UK AI Security Institute, which maintains the open source Inspect framework for LLM safety evaluations.

    \item \textbf{Promote a unified definition of open source AI}. The open LLM models surveyed come under various licensing conditions, complicating reuse and open collaboration of the models and across their communities. Using or requiring licenses that align with the OSI's Open Source AI definition in government-funded projects could clarify usage conditions, promote open innovation, and advance policy goals related to both interoperability and digital sovereignty.
\end{itemize}

\subsubsection{Platform providers}
\begin{itemize}
    \item \textbf{Enable multi-modal contribution pathways beyond code.} Create infrastructure for data annotation, evaluation feedback, documentation, and non-technical contributions. Cohere for AI's success with 3,000+ contributors demonstrates the value of tools that make participation accessible.
\end{itemize}

\subsubsection{Academic institutions}
\begin{itemize}
    \item \textbf{Default to using licenses for open LLMs that uphold the four freedoms of open source, as proposed by the Open Source AI definition \cite{osi_open_2024}.} The use of open and standardised licenses that uphold the four freedoms of open source---use, study, modify, and redistribute---enables collaboration within, beyond, and between research project lifespans, facilitating technology transfer, industrial applications, and advancements across scientific fields and disciplines. For example, the OpenMDW license is the first permissive license specifically designed for AI models, which upholds all four freedoms of open source \cite{OpenMDW2025}.
\end{itemize}

\subsubsection{Open source foundations}
\begin{itemize}
    \item \textbf{Provide support structures for open LLM projects.} Collaboration in open LLM projects across artefacts, such as software, data and models, requires a diverse set of processes, governance models, and tools to function. Open source foundations can provide support structures and neutral grounds for stakeholders across industry, government, and academia to collaborate on the development of open LLMs and related artefacts.
\end{itemize}

\subsection{Threats to Validity}
\noindent  The validity of our research, as well as the threats to it, was informed by following the guidance for qualitative software engineering research~\cite{runeson_guidelines_2008, easterbrook_selecting_2008}.

\subsubsection{Internal Validity}
\noindent  Internal validity concerns the extent to which causal relationships can be established and confounding factors are controlled~\cite{easterbrook_selecting_2008}. We acknowledge several threats to internal validity. First, our purposeful sampling approach, whilst appropriate for exploratory research, may introduce selection bias as we prioritised models with high activity metrics (i.e., downloads and discussions in model repositories on Hugging Face Hub). This could systematically exclude smaller or emerging collaborative efforts that employ different practices and lack significant network reach to publicise their work. To address this concern, we also purposively sampled open LLM projects known to be spearheading open collaboration, such as EleutherAI and the BigScience Workshop. Second, our reliance on developer perspectives may create a single-viewpoint bias, as we did not interview other stakeholders, such as users or contributors, who might offer different insights into collaborative practices. Still, many of our interviewees take on both user and contributor roles in how they reuse extant resources (e.g., datasets) and base models in their work. Third, the retrospective nature of interviews may introduce recall bias, as interviewees' recollections of collaborative processes may be incomplete or influenced by subsequent experiences. Finally, social desirability bias may affect responses, as interviewees may present their collaborative practices in a more positive light than they actually are. The latter threats were addressed through reviewing on online sources about the models and the related collaborations (when possible) to provide contextual awareness to the analysis.

\subsubsection{External Validity}
\noindent  External validity concerns the generalizability of the findings~\cite{easterbrook_selecting_2008}. We acknowledge that our qualitative analysis of 14 open LLM projects limits the generalizability of our findings, and the projects are certainly not representative of all open LLM projects. The sample represents a specific subset of successful and well-known open LLM projects, potentially excluding smaller-scale, failed, or less visible collaborative efforts. Our sample also does not represent emerging models of collaboration such as ``open labs'' (e.g., Marin), which have emerged toward the end of this research project \cite{hall2025marin}. Additionally, our focus on models hosted on the Hugging Face Hub may not capture collaborative practices on other platforms or through other distribution channels. While we acknowledge these limitations, we underscore that our research aim is exploratory, and, towards this end,, our sample provides a diverse enough set to offer a first glimpse into the context of open collaborative development of open LLMs and open source AI at large. 

\subsubsection{Construct Validity}
\noindent  Construct validity concerns the extent to which the measurements accurately represent the phenomenon under study~\cite{easterbrook_selecting_2008}. Our use of download numbers and discussion counts in the model repositories on Hugging Face Hub as proxies for technical and social activity, respectively, may not fully capture the complexity of collaborative engagement. For instance, models with fewer downloads might still involve rich collaborative processes, whilst high download counts might reflect individual rather than collaborative usage. The categorisation of releasing entities into five types (large enterprises, SMEs, public research institutes, non-profit/grassroots, individuals) required subjective interpretation based on available online sources, potentially leading to misclassification. Additionally, our extended model lifecycle framework, whilst grounded in existing literature, represents our conceptual interpretation and may not align with how practitioners themselves understand collaboration boundaries and stages.

\subsubsection{Reliability}
\noindent  Reliability concerns the consistency and replicability of the data collection and analysis procedures~\cite{easterbrook_selecting_2008}. To enhance reliability, all interviews were recorded and transcribed using offline transcription software to ensure accurate data capture. Our data coding and analysis process involved multiple authors using an abductive coding approach, with dual coding for all interviews to reduce potential biases that may arise when a single author performs qualitative data analysis alone~\cite{cruzes_recommended_2011}. Audit trails were maintained throughout the research. Regular peer-debriefing sessions among the author team helped maintain consistency in interpretation and coding. In addition, we conducted member checking by sending the results to respondents for verification, thereby increasing the reliability of the findings~\cite{lincoln_naturalistic_1985}. The iterative development of our codebook, informed by the theoretical framework and refined through joint discussions, provides transparency into our analytical approach. Despite these efforts, we acknowledge that the subjective nature of qualitative coding means that different researchers might interpret the same data differently, and our findings remain inherently influenced by the research team's perspectives and prior knowledge of the open source AI ecosystem.

\section{Conclusion}\label{sec:conc}
\noindent We conclude that open collaboration in open source AI is neither uniform nor static. Opportunities for open collaboration go beyond software and model-centric views, spanning multiple artefact domains, including models, data, software, evaluation, compute, and community engagement. The types and levels of collaboration happening are impacted by differences in accessibility, expertise requirements, modularity, and control. Moreover, collaboration evolves across the LLM development lifecycle, shifting from concentrated, selective engagement in the early stages to broader, distributed participation after model release. We further show how open LLM developers are motivated by a variety of social, economic, and technological motivations, ranging from democratising access to AI and promoting open science to building regional ecosystems and expanding language representation. We also present how collaborative development in open LLM projects can be organised and coordinated through a range of governance structures, typically formal and professionalised to varying degrees, ranging from centralised company-led efforts to decentralised grassroots initiatives. By synthesising the findings into a conceptual model for open LLM collaboration and recommendations for practice, we aim to contribute to research and practice in promoting and enabling greater collaborative development in open LLM projects and open source AI more broadly.


\appendix
\section{Acknowledgments}
\noindent  We would like to thank our research contributors from Ai2, AI Singapore, the BigScience Workshop, BAAI, Cohere Labs, EleutherAI, Hugging Face, Meta, the SpeakLeash Foundation, SCB 10X, Ant Group, the National Library of Norway, Masakhane, and OpenGPT-X for contributing their invaluable time and expertise to this study. In addition, we would like to thank Alek Tarkowski, Andrew Strait, Elizabeth Seger, Felix Sieker, Max Gahntz, Peter Cihon, and Matt White for their constructive feedback on previous versions of this manuscript.

\section{Collaboration Challenges and On-Ramps}
\label{app:collab-challenges-and-onramps}

\noindent Collaboration Challenges (Ch) and Cn-ramps (OnR) per phase (pre-training (PreTr), Post-training (PostTr), Post-release Reuse (PostRe)) and area of collaboration (model, data, software, evaluation, non-technical, compute). Challenges and On-ramps are abbreviated [Phase][Challenge/On-ramp number in the phase].

\subsection{Pre-training: Model Collaborations}

\noindent\textbf{PreTrCh-1: Organisations guard pre-training methods as competitive secrets.}
Many organisations treat information about their model development processes as a competitive advantage, creating barriers to open collaboration. This culture of secrecy particularly affects the sharing of pre- and post-training methodologies, as organisations are reluctant to disclose technical details that could benefit competitors. As I8 from Ai2 explains, \textit{``The modelling part with pre-training and post-training is treated as, like, a very closely guarded secret. We have informal chats, even like on the record, that are somewhat easy to make happen, but formal collaborations, that is a different beast.''}

\noindent\textbf{PreTrCh-2: Resource-intensity limits spontaneous collaboration opportunities.}
The resource-intensive nature of LLM pre-training requires extensive upfront planning that can stifle organic collaboration opportunities. Developers must scope potential collaborations early in the development cycle because late-stage partnerships can fundamentally alter experimental approaches and resource allocation. For example, I8 from Ai2 notes that, \textit{``[Collaborations] all need to be scoped out basically before we start working on something, because they will majorly alter the experimentation stage.''} While some spontaneous collaborations do sometimes emerge through ad hoc interactions at conferences or new hires, these typically remain limited to small-scale tests rather than core model development. Coordination of pre-training activities can be more challenging when organisational technical practices and security constraints on access to resources required for pre-training (e.g., servers, databases) create barriers and thus disincentives for onboarding external collaborators.

\noindent\textbf{PreTrCh-3: Technical complexity restricts participation to experts.}
The technical complexity of LLM pre-training creates natural barriers to broad community participation, effectively limiting meaningful contributions to small teams of domain experts. This expertise bottleneck is particularly pronounced in regions with fewer AI specialists. I10 from the SpeakLeash Foundation highlights this challenge: \textit{``the training is [done] mostly by our internal team. Because one thing is that this is very specialised knowledge and not everyone has it... And currently, I believe we do not have many more experts on this level in Poland.''} Developers recognise that scaling participation beyond core technical teams would be counterproductive, as I10 explains, \textit{``it would be very unoptimized to have like 300 people involved in it.''} The BigScience Workshop similarly organised itself into smaller, expertise-based working groups per domain area. I4 explains, \textit{``one working group focused on the retrieval augmentation experiment, another working group was looking at whether we were going to do a full-length one-directional language model or a sequence-to-sequence language model. Another working group was looking at the hardware, working with the cluster directly.''}

\noindent\textbf{PreTrCh-4: Fast-paced development cycles prevent sustained collaborations.}
The fast-paced nature of open LLM development makes it difficult to sustain long-term engagement and build lasting collaborative relationships in these projects. I14 from Ant Group observes that, \textit{``many AI open source projects are very short-lived. They do not live longer. They just appear for one or two months, and they will disappear forever.''} 

\noindent\textbf{PreTrOnR-1: Collaborative reproduction of existing models for broader access.}
Developers with complementary expertise collaborate to reproduce existing models to address licensing limitations and expand access within the open source community. These reproduction efforts leverage shared codebases and architectural similarities to accelerate development. For example, EleutherAI and Hugging Face collaborated on implementing Meta's first Llama model, with I3 from EleutherAI noting that, this \textit{``Llama implementation in Hugging Face is actually authored by a mixture of Hugging Face staff and EleutherAI volunteers''} and was \textit{``substantially based on our model code because the Llama model and our model are architecturally very similar.''}

\noindent\textbf{PreTrOnR-2: Strategic partnerships between foundation model and specialised model developers.}
Developers of base and derivative models collaborate formally and informally to enhance both model capabilities and regional representation. These collaborations involve information sharing, early access arrangements, and data contributions and partnerships that benefit the broader ecosystem. For example, AI Singapore has engaged with both the Llama team at Meta and the Gemma team at Google, with different levels of collaboration in each case. While the engagement with Meta involved communication through private Slack channels and early access arrangements, the collaboration with Google has been more extensive, including joint development work such as building SEA-LION v4 together. Both collaborations were motivated by the principle that \textit{``it is a net benefit if Southeast Asian languages and cultures are better represented across all models.''}. These engagements have included communication and information sharing on model roadmaps, discussions of specific features, and various forms of technical collaboration. 

\noindent\textbf{PreTrOnR-3: External domain experts contribute specialised knowledge through advisory roles.}
External domain experts contribute specialised knowledge to collaborative projects through advisory roles, which is particularly valuable when core teams lack specific technical expertise. This enables projects to access high-level guidance without requiring contributors to make a full-time commitment. For example, I16 from Masakhane describes how, \textit{``we ask for technical advice. So we did some project on machine translation evaluation. I have not done this before, so we requested at Masakhane, and one professor from the U.S. was interested [and said,] 'Okay, I can help.' She has been doing machine translation for many, many years, so she contributed mostly on technical skills.''}

\subsection{Pre-training: Data Collaborations}
\noindent\textbf{PreTrCh-5: Copyright restrictions and legal complexity constrain dataset sharing.}
Copyright restrictions and rights-holder interventions create significant legal complexity that constrains both dataset creation and model release strategies, leading developers to implement careful filtering and dual-licensing approaches to ensure compliance with the law. I15 from the National Library of Norway explains: \textit{``Recently there has been increased attention from rights holders to clarify what may be used for training and what may be released.''} This pressure prevents the use of permissive licenses such as MIT or Apache v2 when models are trained on copyright-restricted material, forcing developers to consider releasing dual models—one under the Apache license trained exclusively on open-access data. I15 explains that their team routinely consult legal counsel to ensure compliance, while grassroots projects like Masakhane rely on academic and community conventions and use licenses like CC BY 4.0 for commercial or non-commercial use, lacking resources for extensive legal consultation. 

\noindent\textbf{PreTrCh-6: Data curation receives disproportionately less attention than model training.}
Data preparation and curation receive disproportionately less attention and resources compared to model training, creating a systemic imbalance that affects dataset quality and availability. This preference hierarchy reflects broader structural incentives in the AI ecosystem that prioritise visible technical achievements over foundational data work. I15 from the National Library of Norway observes, \textit{``Everybody wants to train a model. Everybody wants to build their expertise on training the models. It is a bit less competitive in the data preparation. To be honest, so nobody wants to do the data work.''} This imbalance not only affects resource allocation but also limits the pool of contributors willing to engage in the time-intensive work of dataset creation and validation.

\noindent\textbf{PreTrCh-7: Reluctance to share datasets due to high investment costs.}
Despite high demand for quality datasets, organisations remain reluctant to share curated data due to the substantial time investment required to prepare them and the perceived competitive value, creating an asymmetric ecosystem where data consumption exceeds contribution. The tedious nature of data curation, requiring hours to produce even small high-quality datasets, compounds this reluctance even when legal permissions exist. I15 from the National Library of Norway describes this challenge: \textit{``Data curation is very tedious work, right? It is very boring work... traditionally, to make a dataset, a fine-tuning dataset, even making 500 or 1000 examples of this, right? It takes hours and hours and hours. And even if you technically are allowed to release it, not everyone is willing to. They want to have it themselves because they invested so much money in it.''} This lack of reciprocity presents a fundamental structural barrier to building robust open data ecosystems.

\noindent\textbf{PreTrOnR-4: Building upon established foundational datasets (e.g., CommonCrawl or The Pile).}
Developers frequently collaborate by building upon established open datasets like CommonCrawl and The Pile, which serve as foundational sources that enable collaboration across the ecosystem. These data sources require different levels of processing, with some providing pre-cleaned data while others need extensive filtering and preparation before they can be used to pre-train a model. I7 from BAAI explains their approach: they acquire training data from \textit{``two primary sources: CommonCrawl and the Pile datasets,''} noting that \textit{``CommonCrawl datasets require extra processing, while the Pile datasets are already pre-cleaned.''} I3 from EleutherAI confirms the former, as they had to build a lifecycle for reverse engineering licensing of data available in Common Crawl. In addition, Hugging Face's SmolLM developers have leveraged open datasets like DCLM from DataComp, which was a collaboration between Toyota Research and several other organisations on English data.

\noindent\textbf{PreTrOnR-5: Contribute specialised datasets to address ecosystem gaps.}
Developers contribute specialised datasets that address gaps in the ecosystem's existing training data, particularly for specific capabilities or domains that lack adequate coverage. These contributions often arise from experimental needs surfaced during model development and subsequently benefit the broader community. I9 from Hugging Face explains their motivation for creating the FineMath dataset: \textit{``We tried to reproduce what DeepSeekMath did. They had a strong math model and built a math dataset for pre-training, but they did not release it. So, in this work, we kind of rebuilt datasets similar to what they have, except that it is open, and now everyone can train on it and get really good math performance.''} Similar efforts include creating high-quality code datasets to address training gaps identified during model development.

\noindent\textbf{PreTrOnR-6: Crowdsourced annotation improves dataset quality through distributed expertise.}
Large-scale community engagement enables organisations to improve dataset quality through distributed annotation and feedback, particularly valuable for multilingual datasets where native speaker expertise is essential. These collaborations leverage crowdsourcing platforms and social media to engage both technical and non-technical contributors. For example, I9 from Hugging Face explains that after they released their multilingual FineWeb 2 dataset, which was a collaboration between the Hugging Face training datasets team and researchers at EPFL, they \textit{``created a space on Hugging Face Hub and invited crowd-sourced contributions via social media''} and received contributions from \textit{``a lot of people who participated from around the world,''} including \textit{``people who did not have ML background,''} who helped with annotation and provided feedback on the quality of the data in their native language. The SpeakLeash Foundation similarly leveraged community engagement, deploying a toxicity assessment survey that has received over 60,000 responses.

\noindent\textbf{PreTrOnR-7: Community projects collect culturally specific data for underrepresented languages.}
Community-driven projects focus on collecting culturally specific and linguistically diverse data that would be missed in direct translations from high-resource languages, addressing representation gaps through targeted collaboration with native speakers. These initiatives require careful coordination to establish shared quality standards and cultural guidelines across diverse contributor communities. The Aya project at Cohere for AI exemplifies this approach, creating datasets through data donations focused on \textit{``culturally sensitive and aware topics, for example, question and answer pairs focused on targeted questions around history, literature, recipes, which would be missed in a direct translation from high-resource language data.''} I5 from Cohere for AI emphasises the importance of native speaker involvement: \textit{``it is really important to get the people [speaking the language] involved to do a good job representing them in these models in the end.''} The potential for such collaborations is particularly strong for medium to low-resource languages where research communities recognise shared challenges. I11 from SCB 10X explains: \textit{``I think building datasets is one of the easiest things for us to collaborate. Because...Thai is like a medium to low resource language. So there are not a lot of high-quality datasets. And many in the research community in Thailand understand this. So everyone's really willing to collaborate to build datasets, whether it's a domain-specific dataset or a medical dataset, things like that. It is very easy for us to find partners to collaborate.''}

\noindent\textbf{PreTrOnR-8: Strategic partnerships enable access to high-quality regional language data.}
Inter-organisational partnerships enable access to high-quality regional data by leveraging complementary capabilities and shared objectives for language representation. These partnerships often extend beyond simple data exchange to include technical collaboration and capacity building. For example, AI Singapore collaborates with Google Research to \textit{``collecting and transcribing voice audio specifically for Southeast Asian languages,''} while also working with regional partners to encourage data release by emphasising social-good incentives. Furthermore, I6 from Meta explains that they work with various entities to provide data in different languages, aiming to diversify and broaden the language support of the Llama models, including partnerships with data providers in India for Indian dialects and languages.

\noindent\textbf{PreTrOnR-9: Universities and companies form data-expertise partnerships.}
Collaborations between companies and academic labs create mutually beneficial arrangements in which companies provide data resources and academic labs contribute research expertise and validation. These partnerships often emerge from complementary needs and capabilities across sectors. For example, I15 from the National Library of Norway describes their role as the primary data contributor to a joint research project with Norwegian universities, explaining, \textit{``none of the others have been very interested in it as a research area''} as universities focus more on technical ML, while, \textit{``They really appreciate that we contribute with the open source datasets especially.''} The library also shares restricted data for research purposes with specific academic institutions under appropriate agreements.

\noindent\textbf{PreTrOnR-10: Community develops shared standards for trusted dataset licensing.}
Developers collaborate to address data provenance and mis-licensing issues by developing shared standards and validation processes that build trust in open training datasets. These efforts require community-wide coordination to establish transparent practices and legal certainty. I3 from EleutherAI explains the challenge: \textit{`` `license' laundering' or `data laundering' is a big issue, meaning when data is downloaded, refined, and uploaded several times, which makes determining the provenance of data difficult to assess.''} To address this, EleutherAI has developed collaborative approaches including reverse-engineering CommonCrawl licensing and creating a \textit{``gold list of organisations''} representing trusted sources with verified licensing and provenance, such as screened YouTube channels from academic, governmental, and non-profit organisations. I3 explains that EleutherAI has put a lot of effort into making open datasets fully vetted for copyright, as well as open LLMs trained on those datasets, which will add value to the open source AI ecosystem, giving researchers and developers greater legal certainty.

\subsection{Pre-training: Software Collaborations}
\noindent\textbf{PreTrCh-8: Technical complexity limits external contributions to core development teams.}
The technical expertise and effort required to develop and maintain LLM training tools and infrastructure limit external contributions to open source frameworks, with most development remaining centralised within core teams. I3 from EleutherAI explains that while EleutherAI's GPT NeoX library has received some external contributions from researchers and developers, \textit{``who are using it professionally, either in academia or in industry,''} such contributions are rare due to expertise requirements. For example, when two volunteers contributed the mixture of experts' architecture, I3 notes, \textit{``Those kinds of contributions are quite rare, partially because the skill set is rare and partially because it is like a huge amount of work if you're not receiving benefit from it. It is a really massive amount of work.''} 

\noindent\textbf{PreTrCh-9: Multiple competing frameworks fragment community effort and collaboration.}
The absence of standardised frameworks for training and evaluating LLMs creates ecosystem fragmentation that leads to duplication of effort across the community. For example, there are several open source frameworks for training and evaluating LLMs. While the diversity of frameworks can serve different use cases, the lack of convergence fragments collaboration. For example, I9 from Hugging Face comments, \textit{``the frameworks we released, we are hoping that everyone can try to adopt them. For example, TRL is really now popular and kind of the go-to choice for post-training models and for Nanotron I think there is also some other frameworks that are good, but I think it would be nice if everyone tries to agree on one framework so that we all contribute to the same features.''} 

\noindent\textbf{PreTrCh-10: Internal tool dependencies limit portability and open source value.}
Developers face practical constraints in open-sourcing software when tools are tightly coupled to internal infrastructure or have limited reusability, creating selective barriers to software sharing. While there is general motivation to open source training tools and infrastructure, practical considerations around value and portability influence release decisions. For example, I8 from Ai2 explains that, \textit{``it is always a goal to open source their software for training infrastructure, unless it has very low value in open sourcing (e.g., specifically tied to their infrastructure or is not widely reusable) in which case they describe it in the model card.''} 

\noindent\textbf{PreTrOnR-11: Open source LLM training frameworks reduce barriers to model training.}
Developers develop open source training frameworks to enable broader community access to LLM training capabilities while benefiting from community feedback and contributions. These frameworks become foundational infrastructure that reduces barriers for other teams to train models. For example, I9 from Hugging Face explains their decision to develop Nanotron: they \textit{used to use a framework for pre-training from Nvidia but opted to develop their open source framework Nanotron for better flexibility and to have in-house knowledge about large-scale pretraining.''} They also build widely-used libraries, including transformers, TRL, the alignment handbook, and lighteval. I7 from BAAI emphasises the critical importance of such frameworks: \textit{``since [a] large model is so complex, without ...any open source code as the base, it will be very challenging for any team to develop the training code by themselves fully from scratch. It is impossible already.''} EleutherAI's GPT-NeoX exemplifies this impact, with I3 describing it as \textit{``the most widely used open source library for training large-scale AI systems in the world.''} In addition, as we discuss below, the development of open source evaluation frameworks also represents a key area of collaboration. 

\noindent\textbf{PreTrOnR-12: Users contribute specialised capabilities developed for internal needs.}
Developers contribute features to existing frameworks when they develop capabilities for their own use cases and choose to share them with the broader community. These contributions often emerge from practical needs and are sometimes facilitated through supervision or support from the framework maintainers. For example, SynthLabs contributed RLHF finetuning support to EleutherAI's GPT-NeoX after initially developing it for their internal needs. I3 from EleutherAI explains, \textit{``They built on top of our library the capability to do this kind of RLHF finetuning, and then came to us and said, 'Hey, we have this internal code. We want to contribute it back to the community. We have benefited from using your library a lot. We want other people to benefit from the modifications that we have made.' And so we partnered with them to help them bring that to the main branch so other people can use it.''} Similarly, I14 from Ant Group describes collaborative relationships with frameworks like SGLang by XAI, where they \textit{``proposed some requirements and gave feedback based on their experience of using the inference framework and SGLang delivered features based on their feedback.''}

\noindent\textbf{PreTrOnR-13: Collaborating on portability across technical environments.}
Developers have collaborated on porting frameworks or adding support for different underlying technologies, enabling adoption across different technical environments. The AI lab at the National Library of Norway has collaborated with Hugging Face to port PyTorch code to JAX to meet their specific technical requirements. I15 explains, \textit{``we did the port of their code to JAX. They had it in PyTorch, and we really needed that code in JAX... That was a lot back and forth working directly on the code with feedback on the exact part of the code and how to do things.''} While their internal code contained dataset-specific features that were unsuitable for inclusion in the general codebase, the core porting work benefited both organisations and the broader community. In addition, developers have collaborated with hardware vendors and compute providers. EleutherAI prioritises portability as a core design principle for GPT-NeoX, with I3 recognising that training is \textit{both a hardware and a software problem.''} I3 goes on to explain that, \textit{``Even if we do not have a use case for a particular supercomputer or a particular context, we want to make sure that it is ready to go in that context before it becomes a requirement for us.''} The portability of GPT-NeoX has been enhanced through contributions from Intel and AMD, who, \textit{``want the framework to run optimally on their accelerators,''} as well as partnerships with national supercomputers like the US Frontier supercomputer and EU-based supercomputers like Lumi and at the University of Barcelona.

\noindent\textbf{PreTrOnR-14: Community identifies and documents fixes for critical software issues}
Information collaborations occur when organisations report and document fixes for bugs identified in core open source software used for model pre-training. I8 from Ai2 describes a case where their team \textit{``found a bug in how numbers have shuffled in PyTorch,''} and reported it to the team. I8 expands that this bug was also discovered by the team at answer.ai when that team \textit{``released their Modern BERT paper... they called this bug explicitly in their tech report and it took I think 48 hours for this bug to be fixed in mainline PyTorch... there is power in like documenting these, even the failures out so that, you know, the other places in the community can react to it and pick them up and fix them.''}

\subsection{Pre-training: Evaluation Collaborations}
\noindent\textbf{PreTrOnR-15: Open source evaluation frameworks enable consistent model comparison across the community.}
The development of evaluation frameworks is an area of collaboration, serving shared community needs for consistent, reproducible model assessment while providing platforms for researchers to contribute new benchmarks and evaluation protocols. EleutherAI's LM Evaluation Harness is an example of this model, enabling comparison of models across evaluation datasets in a \textit{``consistent, reproducible, and reliable''} manner. I3 highlights the value-add of evaluation collaborations: \textit{``if you develop a new evaluation protocol [and] implement it in our library, you're going to get access to all of these models for free, where you do not need to do the work to do the integration yourself.''} The framework's reuse across the community, such as in Hugging Face's OpenLLM Leaderboard, demonstrates its collaborative impact. Hugging Face has built upon this foundation with lighteval, which I9 describes \textit{``complements EleutherAI’s LM Evaluation Harness by adding additional benchmarks and evaluation suites, including Stanford’s HELM.''} I14 from Ant Group describes a similar approach with their ARAIL framework, combining code, datasets, and benchmarks to enable reproducible results: \textit{Our main contribution is actually the code... But to validate the effectiveness of our training framework, we need to run some training and conduct benchmarks. So we also open source the corresponding datasets, so that everyone can reproduce our training results.''}

\noindent\textbf{PreTrOnR-16: Creation of benchmarks for low- and medium-resource languages.}
Evaluation dataset development represents a major collaborative opportunity that brings together diverse partners, including academic institutions, language communities, and model developers, to create comprehensive benchmarking frameworks for specific linguistic or cultural contexts. These collaborations often involve forking existing evaluation platforms and extending them for underrepresented languages and regions. AI Singapore collaborated with Stanford's HELM team to \textit{fork the project and is co-developing an open evaluation framework for Southeast Asian languages.''} Similarly, I12 from Typhoon describes their collaboration on SEA-HELM: \textit{we worked on collecting data for this leaderboard, like Thai exams data, and we basically implemented the code base to make it compatible with the HELM platform,''} while Stanford researchers handled evaluation execution to maintain objectivity. Masakhane has collaborated across multiple evaluation initiatives, including Global MMLU with Cohere for AI, SEACrowd for Southeast Asian contexts, and various African evaluation datasets and leaderboards.

\noindent\textbf{PreTrOnR-17: Speakers of low- or medium-resource languages contribute language expertise to evaluation datasets.}
Language communities and domain experts contribute to the creation of evaluation datasets through crowdsourced efforts that capture cultural and linguistic nuances essential for comprehensive model assessment. These collaborations leverage community knowledge to identify and digitise evaluation materials that would otherwise remain inaccessible. Cohere for AI exemplifies this approach by leveraging language communities as \textit{human eyes''} and \textit{signal''} for understanding model performance across different languages, including campaigns to gather community contributions to global evaluation datasets by finding digitised exam questions from their countries. Similarly, the SpeakLeash Foundation has worked with its community to build datasets for identifying harmful or toxic language, receiving over 60,000 responses and developing specialised benchmarks like CPTU-Bench for complex Polish text understanding and Polish EQ-Bench for Polish emotional intelligence benchmarking.

\noindent\textbf{PreTrOnR-18: Unified leaderboards integrate diverse evaluation resources for comprehensive assessment of LLMs.}
Organisations collaborate by integrating evaluation datasets and benchmarks developed by various research groups into unified leaderboards and evaluation platforms, enabling comprehensive model comparison across diverse capabilities and contexts. This integration work requires technical coordination and shared standards to ensure compatibility and reliability. I12 from Typhoon describes how they \textit{``brought Thai evaluation datasets made by various research groups into a leaderboard and built an evaluation framework for it,''} including building ThaiExam by preparing data and integrating it with the HELM platform to make it available for the broader community. This collaborative approach ensures that evaluation resources developed by different organisations can be leveraged collectively to provide more comprehensive model assessment capabilities.

\subsection{Pre-training: Non-technical Collaborations}
\noindent\textbf{PreTrOnR-19: Shared documentation of training experiences reduces community experimentation costs.}
Developers contribute to the community's collective knowledge base by sharing experimental methodologies, evaluation practices, and detailed training logs that enable others to learn from their experiences and avoid common pitfalls. This knowledge sharing is particularly valuable given the complexity and resource intensity of LLM pre-training, where learning from others' approaches can significantly reduce experimentation costs. I8 from Ai2 emphasises the importance of such sharing: \textit{``Specifically, when it comes to building, the more compute you want to put in your effort, the more design space you can do to achieve your goal, which is like huge, right? And so learning how others have approached similar problems or starting or their final end stage of starting point, or just in general, how they think about problems. It is very valuable.''} They highlight examples including Clementine Fourrier's evaluation work at Hugging Face and the company's detailed blog posts about FineWeb dataset and FineBench, noting, \textit{``It was very interesting to learn about how she thinks about evaluation.''} Such contributions help establish best practices and shared understanding across the community. In addition, developers contribute to this knowledge base by documenting operational challenges, cluster management practices, and detailed training logs, providing invaluable troubleshooting resources for the community. These contributions capture institutional knowledge that would otherwise remain siloed within individual organisations. I8 from Ai2 particularly values comprehensive training documentation, citing examples from major projects: \textit{``training logs from the BLOOM and OPT models by the BigScience Workshop and Meta, respectively... Just like hundreds of pages of everything that can go wrong during pre-training. It was very useful ... you know, software is useful, but there is like this sort of shared knowledge that comes from doing this big project ... in the cases where they get written down, it is actually a super, super useful contribution.''} They also mention the value of blog posts about cluster management practices, which help organisations navigate the complex operational aspects of large-scale training. I8 acknowledges the reciprocal nature of this sharing: \textit{``We tried to pay back all the goodwill that we got in like the almost two technical reports.''}

\noindent\textbf{PreTrOnR-20: Content (e.g., blogs) can amplify a project's visibility.}
Developers collaborate on community outreach efforts to promote awareness of technical contributions, share success stories, and build engagement around collaborative projects. This includes coordinated content creation and marketing efforts that amplify the impact of technical collaborations. The collaboration between EleutherAI and SynthLabs on RLHF finetuning exemplifies this approach, where both organisations published complementary blog posts about the technical contribution. I3 from EleutherAI explains their blog post \textit{``is more focused on the open source ecosystem''} and explains the process of contributing SynthLabs' internal fork upstream into GPT-NeoX. This coordinated approach ensures that collaborative contributions receive appropriate visibility and recognition, educates the community about collaboration processes, and encourages similar contributions.

\subsection{Pre-training: Compute Collaborations}
\noindent\textbf{PreTrCh-11: Compute requirements constrain open collaboration.}
The massive computational requirements for LLM pre-training create fundamental access barriers that limit collaborative model development to organisations with substantial internal resources or partnership arrangements. This resource constraint affects even well-funded organisations and extends beyond raw compute to include the financial and technical complexity of managing large-scale training infrastructure. I3 from EleutherAI emphasises the universal nature of this challenge: \textit{``Resource access is a really big deal. Most organisations do not have on hand the computing resources required to train one of these models. So, figuring out how to make that happen is really important.''} The challenge affects projects at various scales, with I9 from Hugging Face noting that even for smaller models like SmolLM, \textit{``There are significant research and development costs involved not only in running the training, but also finding the right training mixture and parameters.''} Even organisations like BAAI, which have sufficient internal resources for post-training, require external partnerships to access the compute needed for foundation model pre-training.

\noindent\textbf{PreTrOnR-21: Institutional partnerships with public supercomputing infrastructure can enable LLM training.}
National and institutional supercomputing centres actively seek partnerships with open LLM developers to validate their infrastructure while providing essential compute resources that enable large-scale model training projects. These partnerships often emerge from mutual benefits: computing centres need stress-testing and benchmarking, while developers need access to cutting-edge hardware. I4 from Hugging Face explains that BigScience Workshop originated from such a partnership, as \textit{``the origin of the BigScience Workshop was an invitation from an administrator of Jean Zay, a French public cluster, for Hugging Face to stress-test it.''} Over time, this expanded through advocacy efforts to secure additional government investment in GPU infrastructure for the BLOOM model. Similarly, the SpeakLeash Foundation was approached by Poland's Cyfronet supercomputing centre during the construction of the Helios supercomputer. I10 explains, \textit{``the supercomputing centre was currently building the Helios supercomputer, which is currently the most powerful supercomputer in Poland. They got to know about us, and they reached out to us and said, `Hey, we need to run some benchmarks when we are building it. We need to run some synthetic benchmarks but we can also do something for all of useand lend some power to create an LLM,' and we said, `lets do it!'''} In Canada, the national initiative, Compute Canada, to pool computing resources for academic institutions, provides an on-ramp for researchers to access computing resources that would otherwise be prohibitively expensive. I16 from Masakhane explains that, \textit{``everybody pool resources together into what is called Compute Canada, and then you could assess it. Every professor or every AI institute has a quota. Millions of hours that can be used. So if you have really big compute tasks, you can push them to Compute Canada, and then you can run the job.''} They characterise this as providing \textit{``good enough compute, not excellent, but good enough''} for academic-scale projects, while acknowledging limitations compared to industry-scale resources like those needed for models at \textit{``the Llama scale.''}

\noindent\textbf{PreTrOnR-22: Research credit programs enable academic access to enterprise-level infrastructure.}
Major cloud providers offer research credits and specialised programs that enable academic and open source projects to access enterprise-level compute infrastructure, often in exchange for open research publication requirements. These programs provide essential resources for organisations that lack internal computing infrastructure while advancing the providers' research ecosystem engagement. For example, I15 from the National Library of Norway describes receiving substantial resources from Google Research Cloud \textit{``as personal scholarships based on their other work,''} explaining: \textit{``we got quite a lot of resources from them because of the work I've done on vaccine models earlier and the early work we did on BERT. So we train a lot of these original models on TPUs that were given to us. And the only requirement there is really that research needs to be open and we need to publish on it.''} I16 from Masakhane similarly uses compute credits from Google and OpenAI to support their research initiatives, demonstrating how these programs enable resource-constrained academic and grassroots projects to develop open LLMs.

\noindent\textbf{PreTrOnR-23: Vendor partnerships can provide specialised resources and enable technical optimisation.}
Beyond credits, open LLM developers have formed partnerships with hardware and cloud providers to access compute for training LLMs. These partnerships have involved technical collaboration on setup and optimisation in addition to the provision of compute resources. For example, AI Singapore partnered with \textit{``providers like Google, AWS, and NVIDIA, who work with their engineering teams to set up the training lifecycles on the hardware of these providers.''} Similarly, EleutherAI has formed partnerships with chip vendors like Intel and AMD to access specialised hardware configurations and supercomputing resources like the EU's Lumi supercomputer.

\subsection{Post-training Stage}

\subsection{Post-training: Model Collaborations}
\noindent\textbf{PostTrOnR-1: Sharing intermediate checkpoints with trusted partners for testing.}
Some developers share intermediate model checkpoints with selected partners to gather specialised feedback and test specific capabilities during the post-training process, enabling iterative improvement before the model's public release. For example, I9 from Hugging Face explains that they had \textit{``shared the intermediate checkpoint of the model with other startups, so that they could test them and then see if we could add new capabilities to the model during post-training.''} Through these targeted collaborations, they \textit{``got some feedback about the models and then had the final instruct model released,''} demonstrating how selective sharing enables quality improvement through external validation.

\subsection{Post-training: Data Collaborations}
\noindent\textbf{PostTrOnR-2: Sharing specialised post-training datasets with the community.}
Developers share specialised post-training datasets that enable other developers to improve their models. These releases can also have multiplier effects across the ecosystem. For example, I7 from BAAI mentions that they had \textit{``released a post-training dataset called Infinity Instruct, which they had developed for post-training their models like Aquila, on Hugging Face Hub. After three months, it had been used by external developers to post-train over 130 models.''} I7 explains her team is satisfied with the collaboration this facilitates, explaining that, \textit{We open source our dataset, and a lot of people use that to change their model, and they again open source their model. So, we are happy to see this kind of open source cycle.''} 

\subsection{Post-training: Evaluation Collaborations}
\noindent\textbf{PostTrOnR-3: Developers use community benchmark suites for comprehensive model assessment.}
Developers leverage extensive community-developed benchmark suites to conduct thorough model evaluation, recognising that comprehensive assessment requires capabilities beyond those any single organisation can develop independently. This reliance on community resources reflects both practical constraints and best practice recommendations from leading research groups. For example, I9 from Hugging Face describes testing SmolLM \textit{on a very comprehensive list of benchmarks that test really different abilities and also some benchmarks that we did not monitor during the training to make sure ...that we do not overfit on them,''} following practices \textit{recommended by Ai2 in their OLMO paper.''} I7 from BAAI emphasises the necessity of this approach: \textit{if we want to have a full spectrum benchmark for the model,''} it is important to develop benchmarks for various capabilities, but \textit{it is impossible for us to develop all the benchmarks by ourselves.''} Consequently, they \textit{``use a lot of open source benchmarks as part of our evaluation method.''}

\noindent\textbf{PostTrOnR-4: Community leaderboards enable transparent model performance comparisons.}
Organisations participate in community-managed evaluation platforms and leaderboards to facilitate transparent performance comparisons and provide users with standardised benchmarking information. This participation represents a form of indirect collaboration, where organisations contribute evaluation results while benefiting from shared evaluation infrastructure. I7 from BAAI describes their evaluation workflow: after post-training, they \textit{``benchmark their model against other open models. If the results are good, they release the model.''} They also \textit{submit their model to community-managed leaderboards, so that it is easier for the user to compare our model's performance with other third parties' performance.''} This process involves continuous monitoring during training, with developers \textit{``evaluating several checkpoints on a two to three day basis to ensure the curve and trend looks healthy and as expected,''} followed by comparative benchmarking against other open models using community standards and platforms.

\subsection{Post-release Reuse Stage}

\subsection{Post-release: Model Collaborations}
\noindent\textbf{PostReCh-1: Network firewalls create international collaboration barriers.}
Geographical restrictions create friction for international collaboration in open LLM development, manifesting through both network firewalls and discriminatory licensing practices. I14 from Ant Group explains how the firewall in China creates collaboration barriers: \textit{We work internally because there is a firewall. So, we just developed an internal model scope or model states in our internal GitLab, and we upload our updates to GitHub and Hugging Face. We have a regular update, for example, like once a week, and we have some major updates like once every two months or three months.''}

\noindent\textbf{PostReCh-2: Approval-based licensing discriminates against certain regions and organisations.}
LLMs released under restrictive licenses with gated access (i.e., requiring an approval process) create barriers that can discriminate against certain regions or organisational affiliations. For example, I14 from Ant Group describes experiencing this exclusion when attempting to access Llama models: \textit{``I cannot apply for the [Llama] license because my identity says that I'm in mainland China.''} 

\noindent\textbf{PostReOnR-1: Platforms like Hugging Face Hub democratize LLM access.}
Collaboration platforms serve as essential intermediaries that enable widespread model distribution and community access, effectively democratizing model availability beyond direct organisational relationships. These platforms provide standardised infrastructure that reduces barriers to model sharing and discovery. For example, I3 from EleutherAI explains their partnership approach: \textit{``We have been collaborating with Hugging Face on model dissemination... The Hugging Face platform is like the de facto way to share your models and make them accessible to people. And so all of our models are released on the Hugging Face platform.''} Organisations like Masakhane similarly have created dedicated pages on Hugging Face Hub and GitHub where they centralise access to all their datasets and models, facilitating community engagement and reuse.

\noindent\textbf{PostReOnR-2: Government and enterprise partnerships enable customised model deployment.}
Organisations form strategic partnerships with governments and enterprises to enable the adoption of open models for specific use cases, creating collaborative relationships that extend beyond simple model distribution to include technical support and guidance on customisation. These partnerships often serve as showcase examples that demonstrate model capabilities and encourage broader adoption. For example, I6 from Meta describes their government collaborations: \textit{``We are working with the US government with llama and... have made Llama available for their use,''} while highlighting Singapore's advanced adoption: \textit{``One of the most forward-leaning governments I've seen in terms of using these things is the Singapore government. I've talked to a fair amount with their AI team over there, and they deploy open models, including Llama, for a ton of use cases.''} Similarly, I17 from Fraunhofer IAIS describes how Teuken 7B has been adopted by large German technology companies, including IONOS and Deutsche Telekom, with these companies actively promoting the model on social media.

\noindent\textbf{PostReOnR-3: Adapting LLMs for local languages and contexts.}
Regional partnerships have focused on adapting base models to local languages and cultural contexts, often involving multiple stakeholders from a geographic region working toward common goals such as language representation. AI Singapore exemplifies this approach through the SEA-LION model ecosystem, where partners across Southeast Asia have fine-tuned the models for local applications. For example, organisations in Thailand (Visitec) and Indonesia (GoTo) have created derivatives, with GoTo developing a customer service assistant integrated into their payments app that operates in Javanese, Sundanese, and Indonesian. This creates positive feedback loops where \textit{``Many of these partners had contributed data and can now benefit from the open source model output, which further encourages future collaboration and motivation to share and pool resources.''}

\noindent\textbf{PostReOnR-4: Community-driven feedback.}
Developers establish feedback mechanisms that enable community members to contribute feedback about model performance, limitations, and potential improvements. While feedback tends to be reactive rather than proactive, it provides valuable insights for model refinement. I8 from Ai2 acknowledges the value of community input: \textit{``we are getting feedback from the community. That is always useful,''} while expressing interest in more structured engagement similar to the Aya project's approach of \textit{``getting a lot of language help for languages that they do not speak from the community.''} SpeakLeash also proactively collects community feedback through their public-facing interfaces for their Bielik model, where Polish language community members could interact with the model and provide refinement suggestions, with the community ultimately contributing to numerous evaluation benchmarks developed for the project.

\noindent\textbf{PostReOnR-5: Development of derivative models.}
Open model releases enable distributed practitioners to develop derivative LLMs without the developers of the base LLM being directly involved. From finetune to experimentation, these community initiatives often explore directions that core teams do not or cannot pursue due to resource constraints. The Aya project illustrates this dynamic, where following model releases, \textit{``collaborations within the community continue, both using the Aya models and to create further adaptations and derivations. For example, the Aya community built Maya (multimodal Aya), which the core Cohere for AI team was not actively involved in.''} Similarly, I16 from Masakhane describes their involvement in Lugha Llama, a collaborative effort to fine-tune Llama for 20 African languages, which \textit{``outperformed Llama on Global MMLU,''} demonstrating how community adaptations can achieve superior performance for specialised domains.

\noindent\textbf{PostReOnR-6: Creating apps or demos to showcase model capabilities.}
Model developers collaborate with startups and enterprises to create demo applications that showcase model capabilities and explore potential use cases, benefiting both parties through technical validation and market visibility. These partnerships often involve fine-tuning and specialised applications that highlight the versatility of smaller or specialised models. I9 from Hugging Face describes their collaborations with startups around SmolLM: \textit{``These are startups that do really nice things with these small models, so we were able to get some cool demos for our small models.''} They also collaborated with IBM to fine-tune the vision-language version for IBM's DocLing library, demonstrating that \textit{``By specialising small models, you can achieve strong performance, close to that of larger models, while saving significantly on inference costs.''}

\noindent\textbf{PostReOnR-7: Open LLM projects can catalyse follow-up community-driven initiatives.}
LLM releases can catalyse new collaborative initiatives that build upon lessons learned and resources developed, creating successive waves of community-driven projects with improved methodologies and focus. These follow-up projects often inherit infrastructure and knowledge while addressing limitations identified in earlier efforts. For example, I4 explains how the BigScience Workshop catalysed the \textit{``the BigCode project, an open scientific collaboration working on the development and use of LLMs for code, came out of BigScience and was able to learn from what did not work in BigScience, including having more direction from the start and less experimental working groups.''} I4 characterises BigCode as \textit{``the spiritual successor in many ways that had different values, in part because it was in a different medium, so software code.''} Similarly, I17 from Fraunhofer IAIS describes how Teuken 7B's release \textit{``has ignited new collaborations, including with Hessian AI and the University of Bonn,''} while attracting international interest from researchers who have offered to contribute language datasets for improving the model.

\noindent\textbf{PostReOnR-8): Partnerships on domain-specific applications.}
Some organisations have focused on partnerships in specialised domains where model finetuning requires domain expertise. For example, I6 from Meta explains that they have focused on applying Llama to \textit{``specific use cases like legal applications or medical applications.''} 

\subsection{Post-release: Data Collaborations}
\noindent\textbf{PostReOnR-9: LLM releases can catalyse community efforts to expand language coverage.}
Model releases catalyse ongoing community efforts to expand language coverage and improve data representation for underrepresented languages, often extending beyond the original project scope through sustained collaborative initiatives. These efforts leverage the infrastructure and momentum established during initial development while addressing gaps identified through community feedback. For example, I4 from the BigScience Workshop describes how BLOOM's release enabled continued development: \textit{``developers extended BLOOM by adding new languages after most of the projects in the BigScience project had ended.''} This included both technical extensions, such as a developer who \textit{``built a chat demo application from BLOOM for multilingual chat,''} and sustained data collection efforts, where \textit{``an ongoing Arabic data initiative was able to expand its data sourcing by working with the BigScience Workshop,''} with I4 noting that \textit{``a lot of the work happened in parallel with the Arabic data sourcing for BigScience and kept going on.''}

\noindent\textbf{PostReOnR-10: Creation of specialised datasets for emerging use cases.}
Open model releases inspire the development of specialised datasets that address specific use cases or methodological needs identified through model deployment and community experimentation. These derivative datasets often solve technical challenges that emerged during practical applications, contributing valuable resources back to the broader ecosystem. I8 from Ai2 explains how their open models enabled new dataset development: \textit{``datasets and new model releases have emerged from Ai2's open models. This includes the Persona project, which synthetically built attributes and personas used to improve data generation,''} which I8 characterises as having been \textit{``a big unblocker for the team.''} These derivative datasets demonstrate how open model releases can catalyse innovation in data generation methodologies that benefit the broader community.

\noindent\textbf{PostReOnR-11: Creating data collection platforms can enable long-term community engagement.}
Some developers have established data collection mechanisms following model releases to sustain ongoing community engagement and enable continuous improvement of language representation and model capabilities. These infrastructure investments create long-term collaborative platforms that extend the impact of initial model releases. For example, the Cohere for AI team built a dedicated website for continuously collecting language data following their Aya model releases, creating a sustainable mechanism for community members to contribute linguistic and cultural knowledge that can inform future model development and ensure continued expansion of language coverage and cultural representation.

\subsection{Post-release: Software Collaborations}
\noindent\textbf{PostReCh-3: Resource constraints limit the ability to make open sourced software generalizable.}
Developers often lack the resources necessary to transform internal tools into community-ready software, creating barriers to broader adoption despite open source releases. The effort required to generalise, document, and maintain software across diverse use cases extends well beyond the primary research objectives. For example, I15 from the National Library of Norway encountered this challenge with their Whisper model code: \textit{``Absolutely everything is open, but that is more like a principle, right? I think there are two, three, maybe four people who have actually taken that code and used it as a basis for their project... we have not put a lot of effort into making this code generic. Also, we do not really have the resources for that. It is not our role to make general JAX training codes for Whisper out there, right?''} They acknowledge that \textit{``It requires quite a lot to get this up to a level where absolutely everybody can use it and everything is well enough documented,''} highlighting the tension between open principles and practical resource constraints.

\noindent\textbf{PostReOnR-12: Grant-funded research groups contribute significant framework improvements upstream.}
Academic research groups make significant improvements to open source frameworks by adapting tools in their projects, often upstream optimisations and features that benefit the entire community. These collaborations typically emerge from grant-funded research projects that have both the resources and motivation to contribute back to the ecosystem. For example, I3 from EleutherAI describes a substantial collaboration with Mila and the LAION project, which \textit{``had received a grant to train a model on the US national Summit (now Frontier) supercomputer, used EleutherAI's GPT NeoX library to train their models.''} Their work resulted in the Red Pajamas models and dataset, with I3 explaining that \textit{``this adaptation to get [the model] running easily on Summit was contributed upstream [to GPT-NeoX] by them because they wanted other people to be able to do the same things that they were able to do.''} Additional technical contributions have included advanced features like QLoRa implementation by Tim Detmers, \textit{``who invented quantised low rank adaptation, helped to enable Quantised Low-Rank Adaptation (QLoRa) of models allowing them to run in very low resource settings.''}

\noindent\textbf{PostReOnR-13: External users provide testing feedback and performance improvements.}
Developers receive contributions in the form of testing and feedback from external researchers and developers. These contributions often focus on improving usability, performance, and expanding applicability to new use cases. For example, I8 from Ai2 describes feedback on their OlmoOCR PDF parser: \textit{``We got feedback from some folks at different organisations like... the folks in the Eleuther community who ... provide us with interesting test cases to use [and] we have academic collaborators who say... 'well, this could be useful for this set of manuscripts. You should consider whether it can work on transforming this.'''} They also received practical contributions to their dataset analysis tools, including performance improvements from \textit{``a developer working on Korean language models who contributed performance-related data tools, which they ended up reusing themselves.''} I14 from Ant Group emphasises that community feedback primarily focuses on code accessibility: \textit{``There is rarely a question about the reproducibility, but there are many questions about the readability. because you know our code is very engineering focused so it is not that easy to read for researchers... so we made some modifications given their suggestions.''}

\noindent\textbf{PostReOnR-14: Framework adaptation and tool development.}
In some cases, developers develop and release new frameworks when existing tools do not meet their specific requirements, contributing specialised solutions that address gaps in the ecosystem. These contributions often emerge from practical needs during model development and can provide alternatives that serve different use cases or technical environments. For example, I17 from Fraunhofer IAIS describes developing their Modalities LLM framework after finding limitations with existing options: \textit{``we changed it because it was not easy for us to use it out of the box,''} leading them to release their framework under MIT license alongside their Teuken 7B model, providing the community with additional options for LLM development.

\subsection{Post-release: Evaluation Collaborations}
\noindent\textbf{PostReOnR-15: Developers use shared evaluation tools to benchmark and compare LLM capabilities.}
Developers actively participate in community-managed benchmarking platforms that provide standardised evaluation frameworks and transparent model comparison capabilities, benefiting from shared evaluation infrastructure while contributing to ecosystem-wide performance tracking. These platforms serve as neutral ground for model comparison and help users make informed decisions about model selection. For example, I7 from BAAI explains their approach: \textit{``we also submit [our models] onto some open community benchmarks. Usually, these benchmarks are owned by an open source community. We submit our model to them and use their benchmark to get another benchmark result, which will be easier for the users to compare our model's performance with the performance of third parties' models.''} This participation creates mutual value by providing developers with comparative context while enriching community benchmarking resources with additional model coverage. 

\noindent\textbf{PostReOnR-16: Continuous benchmarking.}
Evaluation platforms establish mechanisms for ongoing community contributions that expand benchmark coverage, update evaluation results, and incorporate new models and capabilities as the field evolves. These collaborative platforms create sustainable ecosystems where evaluation resources grow through distributed community effort rather than centralised development. I16 from Masakhane describes their AfroBench initiative: \textit{``recently, we had this project called Afrobench... that tries to pull across all the resources that have been created across Africa. It is a dataset, and then we create a benchmark, and eventually we created a leaderboard.''} The platform actively solicits contributions, with I16 explaining, \textit{``we asked for contributors so as this paper gets out people are providing a lot of feedback and people even contribute new evaluation so for example we have someone say okay i have the results for gpt 4.1 we didn't cover it in our paper they provided all these results... we want to make this open that if you evaluate a new model just give us the results and then we'll put it in the ranking.''} This approach enables continuous expansion and improvement of evaluation capabilities through community engagement.

\subsection{Post-release: Non-technical Collaborations}
\noindent\textbf{PostReOnR-17: Research on open artefacts provides valuable feedback to original developers.}
Researchers use open LLMs to generate valuable insights for base model developers, providing indirect feedback that influences future development decisions. I8 from Ai2 describes the impact of external scientific research that uses open models and facilitates open model development. For example, they highlight the ``Fishing for Magikarp'' paper by Land \& Bartolo~\cite{land_fishing_2024}, who found that unique token sequences (``glitch tokens'') lead to unpredictable but repeatable model failures, which can be at least partly traced to undertrained tokens, a hypothesis that is much easier to verify for models whose pretraining corpus is fully released. I8 notes that this research provided concrete insights about how \textit{``vocabulary may impact stability of a run during training...So, it was not like a direct contribution, but it was a finding that builds specifically on top of our artefacts, and then it formed like a whole line of exploration that made our models better.''} This demonstrates how transparency enables research that creates unexpected value for original developers.

\section{Description of sampled open LLM projects}
\label{app:cases}
\subsection{Allen Institute for AI (Ai2)} 
Ai2 is a US-based non-profit research institute that develops open models and tools to support science and education. Its OLMo language model series (e.g., OLMo 7B, OLMo 1.7 7B, OLMoE, and OLMo 2 7B/13B/32B ) is trained on openly licensed datasets, with transparent training documentation and reproducible code. Ai2 also develops fully open postraining pipelines (Tulu model series) and multimodal vision-language models (Molmo). Ai2 also contributes to developing software tools (e.g, OLMo Core, Open Instruct, OLMES) and benchmarks (e.g., ARC, HellaSwag). Hugging Face: 766 models, 265 datasets. See more: \url{https://allenai.org/} and \url{https://huggingface.co/allenai}.

\subsection{AI Singapore} 
AI Singapore is a government-funded programme housed at the National University of Singapore. It develops open LLMs adapted to Southeast Asian languages and use cases, such as the SEA-LION model series and datasets covering Singlish, Malay, and other local languages. For example, they have released the SEA-PILE and SEA-HELM datasets for pre-training and evaluating LLMs for SEA languages. Its work supports Singapore's national AI strategy, with international collaboration through the Open LLM Leaderboard and multilingual benchmarks. They also have a website (\url{https://sea-lion.ai/}), a playground for developers to chat with their models and create API keys to use their models (\url{https://playground.sea-lion.ai/}), and a beta version of a data hub for SEA data (\url{https://aquarium.sea-lion.ai/}). Hugging Face: 44 models, 18 datasets. See more: \url{https://sea-lion.ai/} and \url{https://huggingface.co/aisingapore}.

\subsection{BigScience Workshop} 
The BigScience Workshop was a large-scale, year-long collaboration coordinated by Hugging Face in 2021–22, involving over 1,000 researchers globally from academia, civil society, and industry. Its main outputs were the BLOOM model, a 176B-parameter multilingual LLM, and the ROOTS dataset, which contains over 60 multilingual corpora. The project also pioneered governance and licensing norms for open LLMs, including the Responsible AI License (RAIL). Hugging Face: 162 models, 10 datasets. See more: \url{https://bigscience.huggingface.co/blog/bloom} and \url{https://huggingface.co/bigscience}.

\subsection{Beijing Academy of AI (BAAI)} 
The Beijing Academy of AI is a Chinese non-profit research institute supported by industry and academic institutions. BAAI has released several open LLMs, including the Aquila, OpenSeek and WuDao families, along with large-scale datasets such as CCI4.0-M2 (a Chinese-English mixed corpus). It forms the large-model open source technical stack, FlagOpen, including open sourced datasets, open sourced benchmarks, open sourced system software FlagOS and open sourced algorithms FlagAI and lots of open sourced models. Hugging Face: 149 models, 111 datasets. See more: \url{https://huggingface.co/BAAI}.

\subsection{Cohere Labs} 
Cohere Labs (formerly Cohere for AI) is a non-profit research lab affiliated with Cohere, a Canadian AI company. It focuses on responsible open science and multilingual research. Key projects include Aya, a family of models including Aya 101, Aya Expanse, and Aya Vision, as well as an open multilingual benchmark for over 200 languages. The group also supports ethical dataset creation and publishes detailed model documentation. Hugging Face: 16 models, 18 datasets. See more: \url{https://cohere.com/research} and \url{https://huggingface.co/CohereLabs}.

\subsection{EleutherAI} 
EleutherAI is a grassroots research collective originally formed in response to OpenAI's decision not to release GPT-3. It has produced several open LLMs, including GPT-J, GPT-NeoX, and the Pythia series, as well as the popular GPT-NeoX training framework. EleutherAI emphasises open access, research transparency, and critical reflection on model risks and limitations, often releasing detailed model cards and training logs. Hugging Face: 669 models, 226 datasets. See more: \url{https://www.eleuther.ai/} and \url{https://huggingface.co/EleutherAI}.

\subsection{Hugging Face} 
Hugging Face is an American startup that operates the Hugging Face platform for hosting, datasets, and tools. In addition to hosting third-party projects, Hugging Face develops its own models (e.g., SmolLM, BLOOM) and OSS libraries, including Transformers, Diffusers, and Evaluate. It plays a central role in convening and coordinating the open source AI ecosystem. The Hugging Face Smol Models Research team has 77 smol models (e.g., SmolLM \& SmolVLM) and 49 training datasets (e.g., FineWeb-Edu, Cosmopedia, Smollm-Corpus). See more: \url{https://huggingface.co/} and \url{https://huggingface.co/HuggingFaceTB}.

\subsection{Meta} 
Meta AI has developed the Llama series of large language models, including Llama 1, Llama 2, Llama 3, and Llama 4. These models are widely used in the open source ecosystem and were released under the Llama Community License Agreement, a commercially permissive license that included an acceptable use policy and attribution requirements. Meta also publishes research on training, evaluation, and alignment, and contributes tooling such as PyTorch, vLLM and open evaluation datasets. Hugging Face: 70 models, 11 datasets. See more: \url{https://www.llama.com/} and \url{https://huggingface.co/meta-llama}.

\subsection{SpeakLeash Foundation} 
SpeakLeash is a Polish non-profit organisation developing open models and datasets for the Polish language, notably the Bielik family of models (v0.1-v3) trained on large-scale Polish corpora collected within SpeakLeash Foundation's core project. They also release quantised Bielik models in order to increase the number of people who can use them on even the least powerful machines. SpeakLeash collaborates with national infrastructure providers and academic institutions to improve Polish AI capabilities. Hugging Face: 67 models, 1 dataset. See more: \url{https://speakleash.org/en/speakleash-a-k-a-spichlerz-english/} and \url{https://huggingface.co/speakleash}.

\subsection{SCB 10X – Typhoon Project} 
SCB 10X is the venture capital and innovation arm of SCBX group, one of Thailand’s largest financial technology business groups. Its Typhoon project aims to develop open source language technologies optimised for Thai. Releases include large language models, vision models, and audio models, along with open, curated Thai datasets. The project supports local AI development and regional innovation. Hugging Face: 54 models, 26 datasets. See more: \url{https://opentyphoon.ai/} and \url{https://huggingface.co/scb10x}.

\subsection{Ant Group} 
Ant Group is an affiliate fintech company of the Chinese technology firm Alibaba. Within Ant Group, there is a team called inclusionAI that works on open source projects, with a focus on AI, including LLMs and Reinforcement Learning (RL). Their \href{https://huggingface.co/inclusionAI}{Hugging Face profile} states that their work \textit{``is guided by the principles of fairness, transparency, and collaboration, and we are dedicated to creating models that reflect the diversity of the world we live in.''} Hugging Face: 28 models, 11 datasets. See more: \url{https://huggingface.co/inclusionAI} and \url{https://github.com/inclusionAI}

\subsection{AI Lab, National Library of Norway} 
The National Library of Norway hosts an AI Lab dedicated to developing open models for Norwegian and other Nordic languages. Projects include NB-BERT and the Norwegian Whisper fine-tuning series, used for speech-to-text transcription of library archives. The lab also releases high-quality datasets in Norwegian and other Nordic languages from the library's archives and other public sector sources. Hugging Face: 166 models, 23 datasets. See more: \url{https://ai.nb.no/} and \url{https://huggingface.co/NbAiLab}.

\subsection{Masakhane} 
Masakhane is a community-led research collective focused on NLP for African languages. It works with researchers, linguists, and communities across Africa and the world to co-create datasets and train models in underrepresented and low-resource languages. Masakhane’s decentralised, multilingual approach includes translation, language modelling, and speech recognition projects. Hugging Face: 359 models, 25 datasets. See more: \url{https://www.masakhane.io/home} and \url{https://huggingface.co/masakhane}.

\subsection{OpenGPT-X} 
OpenGPT-X was a research consortium focused on developing open LLMs that are ``made in Germany,'' involving 10 partners from business, research, and the media. The project was led by Fraunhofer IAIS, a German applied research institute, and funded by the German Ministry of Economic Affairs and Climate Action. The project aimed to build sovereign, open LLMs for European languages and use cases. Its flagship model Teuken 7B is performant in all 24 official EU languages~\cite{ali_teuken-7b-base_2025}. OpenGPT-X has released derivative models, including Teuken-7B-v0.4 and Teuken-7B-v0.6. Hugging Face: 4 models, 0 datasets. See more: \url{https://opengpt-x.de/en/} and \url{https://huggingface.co/openGPT-X}.



\begin{small}
\bibliographystyle{elsarticle-num}  
\bibliography{refs,references-cailean}
\end{small}

\end{document}